\documentclass[pra,aps,showpacs,onecolumn,twoside,superscriptaddress]{revtex4}

\usepackage{amsmath,amsfonts,amssymb,color,epsfig,graphics,graphicx,latexsym,revsymb,theorem,url,verbatim}

\newtheorem{definition}{Definition}
\newtheorem{proposition}[definition]{Proposition}
\newtheorem{lemma}[definition]{Lemma}

\newtheorem{theorem}[definition]{Theorem}
\newtheorem{corollary}[definition]{Corollary}
\newtheorem{conjecture}[definition]{Conjecture}

\newtheorem{remark}[definition]{Remark}
\newtheorem{example}[definition]{Example}

\def\squareforqed{\hbox{\rlap{$\sqcap$}$\sqcup$}}
\def\qed{\ifmmode\squareforqed\else{\unskip\nobreak\hfil
\penalty50\hskip1em\null\nobreak\hfil\squareforqed
\parfillskip=0pt\finalhyphendemerits=0\endgraf}\fi}
\def\endenv{\ifmmode\;\else{\unskip\nobreak\hfil
\penalty50\hskip1em\null\nobreak\hfil\;
\parfillskip=0pt\finalhyphendemerits=0\endgraf}\fi}
\newenvironment{proof}{\noindent \textbf{{Proof.~} }}{\qed}
\def\Dbar{\leavevmode\lower.6ex\hbox to 0pt
{\hskip-.23ex\accent"16\hss}D}
\makeatletter
\def\url@leostyle{%
  \@ifundefined{selectfont}{\def\UrlFont{\sf}}{\def\UrlFont{\small\ttfamily}}}
\makeatother
\def\bcj{\begin{conjecture}}
\def\ecj{\end{conjecture}}
\def\bcr{\begin{corollary}}
\def\ecr{\end{corollary}}
\def\bd{\begin{definition}}
\def\ed{\end{definition}}
\def\bea{\begin{eqnarray}}
\def\eea{\end{eqnarray}}
\def\bem{\begin{enumerate}}
\def\eem{\end{enumerate}}
\def\bex{\begin{example}}
\def\eex{\end{example}}
\def\bim{\begin{itemize}}
\def\eim{\end{itemize}}
\def\bl{\begin{lemma}}
\def\el{\end{lemma}}
\def\bpf{\begin{proof}}
\def\epf{\end{proof}}
\def\bpp{\begin{proposition}}
\def\epp{\end{proposition}}
\def\br{\begin{remark}}
\def\er{\end{remark}}
\def\bt{\begin{theorem}}
\def\et{\end{theorem}}

\newcommand{\nc}{\newcommand}

\def\a{\alpha}
\def\b{\beta}
\def\g{\gamma}
\def\d{\delta}
\def\e{\epsilon}
\def\ve{\varepsilon}

\def\t{\theta}

\def\x{\xi}
\def\p{\pi}
\def\r{\rho}
\def\s{\sigma}

\def\ph{\varphi}
\def\c{\chi}
\def\ps{\psi}
\def\o{\omega}

\def\G{\Gamma}

\def\S{\Sigma}

\def\Ps{\Psi}

\nc{\bbC}{{\mathbb{C}}}

\nc{\cA}{{\cal A}} \nc{\cB}{{\cal B}} \nc{\cC}{{\cal C}}
\nc{\cD}{{\cal D}} \nc{\cE}{{\cal E}} \nc{\cF}{{\cal F}}
\nc{\cG}{{\cal G}} \nc{\cH}{{\cal H}} \nc{\cI}{{\cal I}}
\nc{\cJ}{{\cal J}} \nc{\cK}{{\cal K}} \nc{\cL}{{\cal L}}
\nc{\cM}{{\cal M}} \nc{\cN}{{\cal N}} \nc{\cO}{{\cal O}}
\nc{\cP}{{\cal P}} \nc{\cR}{{\cal R}} \nc{\cS}{{\cal S}}
\nc{\cT}{{\cal T}} \nc{\cU}{{\cal U}} \nc{\cV}{{\cal V}}
\nc{\cW}{{\cal W}} \nc{\cX}{{\cal X}} \nc{\cZ}{{\cal Z}}

\nc{\hA}{{\hat{A}}} \nc{\hB}{{\hat{B}}} \nc{\hC}{{\hat{C}}}
\nc{\hD}{{\hat{D}}} \nc{\hE}{{\hat{E}}} \nc{\hF}{{\hat{F}}}
\nc{\hG}{{\hat{G}}} \nc{\hH}{{\hat{H}}} \nc{\hI}{{\hat{I}}}
\nc{\hJ}{{\hat{J}}} \nc{\hK}{{\hat{K}}} \nc{\hL}{{\hat{L}}}
\nc{\hM}{{\hat{M}}} \nc{\hN}{{\hat{N}}} \nc{\hO}{{\hat{O}}}
\nc{\hP}{{\hat{P}}} \nc{\hR}{{\hat{R}}} \nc{\hS}{{\hat{S}}}
\nc{\hT}{{\hat{T}}} \nc{\hU}{{\hat{U}}} \nc{\hV}{{\hat{V}}}
\nc{\hW}{{\hat{W}}} \nc{\hX}{{\hat{X}}} \nc{\hZ}{{\hat{Z}}}

\def\diag{\mathop{\rm diag}}
\def\dim{\mathop{\rm Dim}}

\def\lin{\mathop{\rm span}}

\def\max{\mathop{\rm max}}
\def\min{\mathop{\rm min}}

\def\rank{\mathop{\rm rank}}

\def\tr{\mathop{\rm Tr}}

\def\GL{{\mbox{\rm GL}}}

\newcommand{\tbc}{\red{TO BE CONTINUED}}

\def\bigox{\bigotimes}
\def\dg{\dagger}
\def\es{\emptyset}

\def\op{\oplus}
\def\ox{\otimes}

\def\sue{\subseteq}
\def\sm{\setminus}

\newcommand{\bra}[1]{\langle#1|}
\newcommand{\ket}[1]{|#1\rangle}
\newcommand{\proj}[1]{| #1\rangle\!\langle #1 |}
\newcommand{\ketbra}[2]{|#1\rangle\!\langle#2|}
\newcommand{\braket}[2]{\langle#1|#2\rangle}

\newcommand{\abs}[1]{|#1|}

\newcommand{\red}{\textcolor{red}}

\newcommand{\ieee}{IEEE. Trans. Inf. Theory}
\newcommand{\jmp}{J. Math. Phys.}
\newcommand{\jpa}{J. Phys. A}

\newcommand{\pla}{Phys. Lett. A}

\def\bZ{{\mbox{\bf Z}}}
\def\bC{{\mbox{\bf C}}}

\begin{document}
\title{Separability problem for multipartite states of rank at most four}

\author{Lin Chen}
\affiliation{Department of Pure Mathematics and Institute for
Quantum Computing, University of Waterloo, Waterloo, Ontario, N2L
3G1, Canada} \affiliation{Centre for Quantum Technologies, National
University of Singapore, 3 Science Drive 2, Singapore 117542}
\email{cqtcl@nus.edu.sg (Corresponding~Author)}

\def\Dbar{\leavevmode\lower.6ex\hbox to 0pt
{\hskip-.23ex\accent"16\hss}D}
\author {{ Dragomir {\v{Z} \Dbar}okovi{\'c}}}

\affiliation{Department of Pure Mathematics and Institute for
Quantum Computing, University of Waterloo, Waterloo, Ontario, N2L
3G1, Canada} \email{djokovic@uwaterloo.ca}

\begin{abstract}
One of the most important problems in quantum information is the
separability problem, which asks whether a given quantum state is
separable. We investigate multipartite states of rank at most four which are PPT (i.e., all their partial transposes are positive semidefinite). We show that any PPT state of rank two or three is separable and has length at most four. For separable states of rank four, we show that they have length at most six. It is six only for some qubit-qutrit or multiqubit states. It turns out that any PPT entangled state of rank four is necessarily supported on a $3\ox3$ or a $2\ox2\ox2$ subsystem. We obtain a very simple criterion for the separability problem of
the PPT states of rank at most four: such a state is entangled if and only if its range contains no product vectors. This criterion can be easily applied since a four-dimensional subspace in the
$3\ox3$ or $2\ox2\ox2$ system contains a product vector if and only if its Pl\"{u}cker coordinates satisfy a homogeneous polynomial equation (the Chow form of the corresponding Segre variety). We have computed an explicit determinantal expression for the Chow form in the former case, while such expression was already known in the latter case.
\end{abstract}

\date{ \today }




\maketitle

\tableofcontents


%
%
%
%
%
%

\section{\label{sec:introduction} Introduction}

In quantum physics, the condition of the spatially distributed
systems is described by multipartite quantum states. The systems can
be uncorrelated, and the corresponding state is \textit{separable}.
Otherwise, the state is entangled \cite{werner89}. Entangle states
are the basic ingredients of quantum-information tasks \cite{bbc93}.
The so-called \textit{separablity problem}, namely to distinguish
multipartite separable states from entangled states, is then a basic
task in quantum information. In this paper, we will address this
problem by studying multipartite states $\r$ of rank at most four, and give a complete answer in this case. We recall that 
if $\r$ is separable then it must be PPT i.e., {\em all partial 
transposes of $\r$ are positive semidefinite}. 
The following separability criterion is based on Lemma \ref{le:PPTranktwo} and Theorems \ref{thm:PPTrankthree},
\ref{thm:PPTrankfour} and \ref{thm:PPTrankfour=noPRODsate}:

\textit{Any multipartite state of rank less than four is 
separable if and only if it is PPT. Any multipartite state $\r$ 
of rank four is separable if and only if (1) $\r$ is PPT, and (2) the range of $\r$ contains a product vector.}

We will show below that this criterion is easy to apply in practice. By the Peres-Horodecki criterion
\cite{hhh96,horodecki97}, the condition (2) above is automatically satisfied by qubit-qubit and qubit-qutrit PPT states of rank four. So the two criteria are equivalent for these states. Moreover, we will show that condition (2) is satisfied by all PPT states of rank four, except for
some states supported on $3\ox3$ or $2\ox2\ox2$ systems. For these states, the violation of condition (2) becomes an essential
tool for the construction of PPT entangled states of rank four.

In general, we consider a quantum system consisting of $n$ parties
$A_1,\ldots,A_n$ with Hilbert space $\cH=\cH_1\ox\cdots\ox\cH_n$. We
denote the dimension of $\cH_i$ by $d_i$ and the dimension of $\cH$
by $d:=d_1d_2\cdots d_n$. We denote by $\G_i$ the partial
transposition operator on the system $A_i$ computed in some fixed
o.n. basis. Thus, if $\r=\r_1\ox\cdots\ox\r_i\ox\cdots\ox\r_n$ then
$\r^{\G_i}=\r_1\ox\cdots\ox\r_i^T\ox\cdots\ox\r_n$, where the
exponent T denotes transposition. We denote by $\G$ the group
generated by the pairwise commuting involutory operators $\G_i$. The
elements of $\G$ are the products $\G_S=\prod_{i\in S}\G_i$ where
$S$ runs through all subsets of $\{1,2,\ldots,n\}$. We say that a
state $\r$ on $\cH$ has PPT if $\r^{\G_S}\ge0$ for all subsets $S$.
Evidently if $\r$ is PPT then so is $\r^{\G_S}$ for any $S$. If a
state $\r$ is not PPT, we shall say that it is NPT. The range and
the rank of any linear operator $\r$ will be denoted by $\cR(\r)$
and $r(\r)$, respectively. Unless stated otherwise, the states will
not be normalized.

A {\em product vector} is a nonzero vector of the form
$\ket{\psi_1}\ox\cdots\ox\ket{\psi_n}$, which will be also written as $\ket{\psi_1,\ldots,\psi_n}$. We say that a subspace of $\cH$ is {\em completely entangled} (CES) if it contains no product vectors.
For counting purposes we do not distinguish product vectors which
are scalar  multiples of each other. A (non-normalized) {\em pure
product state} is the tensor product
$\proj{\psi_1}\ox\cdots\ox\proj{\psi_n}$, where
$\ket{\psi_i}\in\cH_i$ are nonzero vectors. A state is {\em
separable} if it is a finite sum of pure product states. (Some
authors refer to these states as {\em fully separable}.) The {\em
length}, $L(\r)$, of a separable state $\r$ is the minimal number of pure product states over all such decompositions of $\r$
\cite{dtt00}. A state is {\em entangled} if it is not separable. It is immediate from the definition of separable states that every separable state $\r$ is PPT. In other words, an NPT state must be entangled. So it is sufficient to consider PPT states
when we consider the separability problem.

The separability problem has a complete answer for the pure state
$\ket{\ps}$, whose density matrix $\r=\proj{\ps}$ has rank one. By definition $\ket{\ps}$ is separable if and only if it is a product vector. This is equivalent to the statement that any single-party reduced density operator of $\r$ has rank one, which can be easily
verified. Nevertheless, the separability problem becomes quickly
intractable as the dimension and number of systems increase
\cite{gurvits03}. As far as we know, there is no complete answer for multipartite states of rank two despite of some partial results for tripartite systems \cite{abl01,kl01,fgw03}.

In Lemma \ref{le:PPTranktwo} and Theorem \ref{thm:PPTrankthree}, we show that any PPT state of rank two or three is separable. In
Theorem \ref{thm:PPTrankfour}, we show that any PPT entangled state (PPTES) of rank four is necessarily supported on a $3\ox3$ or a $2\ox2\ox2$ subsystem. This is based on the preliminary results including Lemmas \ref{le:3x3x3PPT}, \ref{le:2x3x3,2x3x2PPT} and \ref{le:m(n-m)PPT}. We deduce that any
$n$-partite PPT state with $n\ge4$ and rank four is separable, see Corollary \ref{cr:4partite}. Recall that two-qutrit PPTES of rank four have been recently fully described, see
\cite{bdm99,cd11JMP,skowronek11,cd12JMP}, by using the two-qutrit unextendible product bases (UPB). In particular, the range
of such a state is a CES. We obtain a similar result for
three-qubits. This is based on Propositions
\ref{pp:PPTESrankAB=4} and \ref{pp:3qubitPPTES,property}, which
give various properties of three-qubit PPTES $\r$ of rank
four. For example, Proposition \ref{pp:3qubitPPTES,property} (ii)
states that $\cR(\r)$ is a CES, and (v) states that up to a scalar multiple there is only one PPTES with specified range,
$\cR(\r)$. It is known that both of these properties hold
for two-qutrit PPTES of rank four \cite{cd11JMP}, see below
Theorem \ref{thm:3x3PPTstates} (i),(iv). As a corollary of these results, any PPT state of rank at most four is entangled if and only if $\cR(\r)$ is a CES, see Theorem \ref{thm:PPTrankfour=noPRODsate}. This gives a complete answer to the separability problem for multipartite states of rank at most
four. We also construct some criteria for deciding the
separability of certain states of arbitrary dimensions in Lemmas
\ref{le:1(n-1)PPT}, \ref{le:reducible} and \ref{le:rank<=r}.

We emphasize that Theorem \ref{thm:PPTrankfour=noPRODsate} can be
easily applied to any PPT state of rank at most four. By Lemma
\ref{le:PPTranktwo}, Theorem \ref{thm:PPTrankthree} and
\ref{thm:PPTrankfour}, it is sufficient to consider the two-qutrit
or three-qubit PPT state of rank four, otherwise such state is
separable. It is known that a four-dimensional subspace in the
$3\ox3$ or $2\ox2\ox2$ system contains a product vector if and only
if its Pl\"{u}cker coordinates satisfy a homogeneous polynomial
equation, known as the Chow form of the corresponding Segre variety.
We have computed an explicit determinantal expression for the Chow
form of the $3\ox3$ system, see Eq. \eqref{eq:Fza3x3}, and we also
give the known Chow form for the $2\ox2\ox2$ system in Eq.
\eqref{eq:Fza2x2x2}. They vanish on the range of the two-qutrit or
three-qubit PPT state of rank four if and only if the state is
separable. This is stated in Theorem \ref{thm:3x3stanje4}. The
verification of Eqs. \eqref{eq:Fza3x3} and \eqref{eq:Fza2x2x2} can
be readily performed by an ordinary computer. Note that our method
is analytically operational. It is different from the numerical test
employing methods of semi-definite programming and optimization
\cite{dps04}. As a generalization, we also compute the Chow form of
the Segre variety $\cP^{M-1}\times\cP^1$ in Proposition
\ref{pp:ChowMx2}.

Let us mention some applications of our results. First, it is known that any PPT state is a sum of extreme PPT states, so it is
important to characterize these extreme states \cite{cd12,cd12JMP}. By Lemma \ref{le:PPTranktwo} and Theorem \ref{thm:PPTrankthree}, the PPT states of rank two or three are not extreme. By Theorem \ref{thm:PPTrankfour} and Proposition
\ref{pp:stronglyextreme}, there are only three types of multipartite extreme PPT states of rank at most four: pure product states, two-qutrit PPTES, and three-qubit PPTES. The last assertion coincides with the recent numerical test as reported in \cite{gim12}. Furthermore, it is known that some three-qubit PPTES of rank four can be constructed by using three-qubit UPB \cite{DiV03}. So such states may be related to some Bell inequalities with no quantum violation, which can also be
constructed by three-qubit UPB \cite{aaa12}. We also show that
any PPTES of rank four is strongly extreme, see Proposition
\ref{pp:stronglyextreme}. The latter are extreme PPTES
whose range does not contain the range of any other PPT state
\cite{cd12}, see Definition \ref{def:StronglyExtreme}.

Second, the length of a separable state $\r$ represents the minimal physical effort needed to realize $\r$ by the entanglement of formation \cite{bds96}. Two separable states of  different lengths are not equivalent under stochastic local operations and classical communications (SLOCC) \cite{dvc00}. In Lemma \ref{le:PPTranktwo} and Theorem \ref{thm:PPTrankthree}, besides the results on separability, we also prove that the multipartite separable states of rank two have length two, and
those of rank three have length three or four. We further show that separable states of rank four have length at most six, see
Lemma \ref{le:sepRANK4length}. The bound six is reached by some
known examples of qubit-qutrit separable states, see
\cite[Table 2]{cd12PRA}, as well as by some three-qubit separable states, see the proof of Lemma \ref{le:sepRANK4length} (iii).
These results on lengths of separable states generalize some of the recent results obtained in
\cite{as10,cd12paper8,cd12PRA,hk12,hk12dec}. We may conclude that
the bigger the rank is, the more different lengths the separable states could have.


Third, we present another interpretation of our result in quantum
information. Recall that a  pure state is genuinely entangled if
it is not a product state for any bipartition of the systems
\cite{ss02}. We say that a mixed multipartite state $\r$ is
genuinely entangled if for any ensemble $\{p_i,\ket{\ps_i}\}$
realizing this state, i.e., such that $\r=\sum_i p_i\proj{\ps_i}$,
at least one state $\ket{\ps_j}$ is genuinely entangled. Genuine
entanglement can be detected by many kinds of entanglement witnesses
based on Bell inequalities \cite{ss02} or stabilizer theory
\cite{tg05}. Proposition \ref{pp:3qubitPPTES,property} and Theorem \ref{thm:PPTrankfour=noPRODsate} imply that the multipartite PPT state $\r$ of rank four contains genuine entanglement if and only if it is a bipartite state. In particular, any pure state of the ensemble generating $\r$ is entangled by Theorem \ref{thm:3x3PPTstates} (i). In this sense, genuine PPTES of rank four are rare and not easily prepared quantum resource.

Fourth, if we regard the state $\r$ in Lemma \ref{le:1(n-1)PPT} (i) as a bipartite state of system $A_1$ and all other 
systems $A_2\cdots A_n$ together, then $\r$ is a ``generalized classical'' state as defined in \cite[Definition 1]{ccm11}. In the same paper, analytical methods using algorithms
and a physical criterion have been given for detection of 
generalized classical multipartite states. In Lemma
\ref{le:1(n-1)PPT} (i), we give another method to detect generalized classical PPT states. Let us also mention that the generalized classical states allow for the quantum-information tasks with non-disruptive local state identification. This is related to the so-called local broadcasting \cite{phh08}.

The paper is organized as follows. In Sec. \ref{sec:preliminary} we state the known facts used throughout this paper. In Sec.
\ref{sec:PPTrank23} we study multipartite PPT states of rank two and three. The main results are presented in Lemma \ref{le:PPTranktwo} and Theorem \ref{thm:PPTrankthree}. In Sec. \ref{sec:PPTrank4} we study multipartite PPT states of rank four. The main result is stated in Lemma \ref{le:sepRANK4length} and Theorem \ref{thm:PPTrankfour}. In Sec. \ref{sec:3qubit} we study the three-qubit PPT states of rank four. Based on this result we can characterize all PPTES of rank four. The main results are presented in Proposition \ref{pp:3qubitPPTES,property} and Theorem \ref{thm:PPTrankfour=noPRODsate}. In Sec. \ref{sec:chowform}, we show how one can verify the CES condition for the separability of multipartite states of rank four by using the Chow form. The main result is stated in Theorem \ref{thm:3x3stanje4}.

\section{\label{sec:preliminary} Preliminaries}

From now on, for a given multipartite state $\r$, we denote by $\r_{i_1,\ldots,i_k}$ the reduced density operator on the systems
${A_{i_1},\ldots,A_{i_k}}$.
For any PPT state $\r$ on $\cH$ and any pure state
$\ket{\Ps}\in\ox_{i\in S}\cH_i$, the state $\bra{\Ps}\r\ket{\Ps}$
is a PPT state on the space $\ox_{i\in S'}\cH_i$, where $S'$ is the complement of $S$ in $\{1,2,\ldots,n\}$. Another useful property of multipartite PPT states is that they remain PPT when several local systems are combined into one system. This property enables us to consider multipartite states as bipartite states, and hence simplify many proofs. See for instance, the application of Lemma \ref{le:length=Airreducible} in the proof of Lemma \ref{le:1(n-1)PPT}.


We say that a multipartite state $\r$ is a $r_1\times r_2\times
\cdots\times r_n$ {\em state} if its local ranks are
$r_1,r_2,\ldots,r_n$, i.e., $r(\r_i)=r_i$ for each $i$. For any state $\r$ on $\cH$ and any subset $S\subseteq\{1,\ldots,n\}$, we have
 \bea
 \label{ea:rhoBGamma=rhoB}
 \left( \r^{\G_S} \right)_{i} &=& \left\{
\begin{array}{c}
\r_i^T,\quad i\in S; \\
\r_i,\quad i\notin S.
\end{array} \right.
 \eea
 Consequently,
 \bea
r \left((\r^{\G_S})_{i}\right)=r(\r_{i}),\quad i=1,\ldots, n,
 \eea
i.e., the rank of any single system reduced state is invariant under all partial transpositions.
If $\r$ is an $r_1\times r_2\times\cdots\times r_n$ state, then
$\r^{\G_S}$ is too. If $\r$ is a PPTES so is $\r^{\G_S}$, but they may have different ranks.

As the bipartite case occurs often, we shall use the following
simplified notation: $M=d_1$, $N=d_2$, $A=A_1$, $B=A_2$. When $\r$
is a bipartite state, we refer to the ordered pair
$(r(\r),r(\r^\G_1))$ as the {\em birank} of $\r$. The birank has
been used to characterize many bipartite PPT states. For example,
two-qubit and qubit-qutrit separable states have been classified
independently in terms of the birank in \cite[Table I,II]{cd12PRA}
and \cite{hk04,ck12}, respectively. In the proof of Lemma
\ref{le:reducible} (ii), we will use the fact that the length of a
separable bipartite state is not smaller than the maximum of $r(\r)$
and $r(\r^\G_1)$.

Let us now recall some basic results from quantum information
regarding the separability and PPT properties of bipartite states.
Let us start with the basic definition.

 \bd
 We say that two $n$-partite states $\r$ and $\s$ are
{\em equivalent under SLOCC} {\em (SLOCC-equivalent} or just {\em
equivalent)} if there exists an invertible local operator (ILO)
$A=\bigox^n_{i=1} A_i \in
\GL:=\GL_{d_1}(\bC)\times\cdots\times\GL_{d_n}(\bC)$ such that
$\r=A\s A^\dg$ \cite{dvc00}.
 \ed

It is easy to see that any ILO transforms PPT, entangled, or
separable state into the same kind of states. We shall often use
ILOs to simplify the density matrices of states. For example, any $M$ linearly independent vectors in $\bC^M$ can be converted into the o. n. basis $\{\ket{i}_A:i=0,\ldots,M-1\}$ by an ILO.

Let us extend the formal definition of the term ``general position'' \cite[Definition 7]{cd12} to the multipartite case. 

 \bd \label{def:GenPos}
We say that an $m$-tuple of vectors $(v_1,\ldots,v_m)$ in a vector space $V$ is in {\em general position} if, for any subset
$I\sue\{1,\ldots,m\}$ with $|I|\le\dim V$, the vectors $v_i$,
$i\in I$, are linearly independent. We say that an $m$-tuple
$(\ket{\phi_{k,1},\ldots,\phi_{k,n}})_{k=1}^m$ of product vectors in $\cH=\cH_1\ox\cdots\ox\cH_n$ is in {\em general position} if the $m$-tuple $(\ket{\phi_{k,j}})_{k=1}^m$
is in general position in $\cH_j$ for each $j=1,\ldots,n$.
 \ed

Let us recall from \cite[Theorem 22]{cd11JPA} and \cite[Theorems
17,22,24]{cd11JMP} the main facts about the $3\times3$ PPT states of
rank four. Let $M=N=3$ and let $\cU$ denote the set of unextendible
product bases in $\cH=\cH_A\ox\cH_B$. For $\{\psi\}\in\cU$ we denote
by $\Pi\{\psi\}$ the normalized state $(1/4)P$, where $P$ is the
orthogonal projector onto $\{\psi\}^\perp$.

 \bt
\label{thm:3x3PPTstates} $(M=N=3)$ For a $3\times3$ PPT
state $\r$ of rank four, the following assertions hold.

(i) $\r$ is entangled if and only if $\cR(\r)$ is a CES.

(ii) If $\r$ is separable, then it is either the sum of four
pure product states or the sum of a pure product state and a
$2\times2$ separable state of rank three.

(iii) If $\r$ is entangled, then

\quad (a) $\r$ is extreme;

\quad (b) $r (\r^\G_1)=4$;

\quad (c) $\r\propto A\ox B~ \Pi\{\psi\}~ A^\dag\ox B^\dag$ for some
$A,B\in\GL_3(\bC)$ and some $\{\psi\}\in\cU$;

\quad (d) $\ker\r$ contains exactly 6 product vectors, and these
vectors are in general position.

(iv) If the normalized states $\r$ and $\r'$ are $3\times3$ PPTES of
rank four with the same range, then $\r=\r'$.
 \et
In Proposition \ref{pp:3qubitPPTES,property}, we will show that
the three-qubit PPTES of rank four have similar properties as
$3\times3$ PPTES of rank four. These properties turn out to be
essential in proving the extremality of three-qubit PPTES of rank
four, see Proposition \ref{pp:stronglyextreme}.

From \cite[Theorem 1]{hst99} we have
 \bt
\label{thm:PPTMxNrank<M,N} The $M\times N$ states of rank less than
$M$ or $N$ are distillable, and consequently they are NPT.
 \et

The next result follows from \cite[Theorem 10]{cd11JPA},
\cite{hlv00} and Theorem \ref{thm:PPTMxNrank<M,N}, see also
\cite[Proposition 6 (ii)]{cd11JPA}.

 \bpp
 \label{prop:PPTMxNrankN}
Let $\r$ be an $M\times N$ state of rank $N$.

(i) If $\r$ is PPT, then it is a sum of $N$ pure product states.
Consequently, $r(\s)>\max(r(\s_A),r(\s_B))$ for any PPTES $\s$, any
bipartite PPT state of rank $\le3$ is separable, and the bipartite
PPTES of rank four must be supported on $3\ox3$.
%

(ii) If $\r$ is NPT, then it is distillable.

 \epp

By Theorem \ref{thm:PPTMxNrank<M,N}, in case (i) we must have $M\le
N$. We will generalize the last-but-one assertion of (i) to
multipartite PPT states in Lemma \ref{le:PPTranktwo} and Theorem
\ref{thm:PPTrankthree}. The final assertion of (i) will be
generalized to three-qubit case in Theorem \ref{thm:PPTrankfour}.
The proof of these results require the notion of reducible and
irreducible states constructed in \cite[Definition 11]{cd11JPA}.

 \bd
\label{def:reducibleirreducible} We say that a linear operator
$\r:\cH\to\cH$  is an {\em A-direct sum} of linear operators
$\a:\cH\to\cH$ and $\b:\cH\to\cH$, and we write $\r=\a\oplus_A\b$,
if $\cR(\r_A)=\cR(\a_A)\oplus\cR(\b_A)$. A bipartite state $\r$
is {\em A-reducible} if it is an A-direct sum of two states;
otherwise $\r$ is {\em A-irreducible}. One defines similarly the
{\em B-direct sum} $\r=\a\oplus_B\b$, the {\em B-reducible} and the
{\em B-irreducible} states. We say that a state $\r$ is {\em
reducible} if it is either A or B-reducible. We say that $\r$ is
{\em irreducible} if it is not reducible. We write $\r=\a\oplus\b$
if $\r=\a\oplus_A\b$ and $\r=\a\oplus_B\b$, and in that case we say
that $\r$ is a {\em direct sum} of $\a$ and $\b$.
 \ed

The next result on reducible states is from \cite[Lemma 15]{cd12}.

\bl
 \label{le:rhoreducible=rhoPTreducible}
Let $\a$ and $\b$ be linear operators on $\cH$.

(i) If $\r=\a\oplus_B\b$, then $\r^\G_1=\a^\G_1\oplus_B\b^\G_1$.

(ii) If $\a$ and $\b$ are Hermitian and $\r=\a\oplus_A\b$, then
$\r^\G_1=\a^\G_1\oplus_A\b^\G_1$.

(iii) If a PPT state $\r$ is reducible, then so is $\r^\G_1$.
 \el

Let us recall a related result \cite[Corollary 16]{cd11JPA}. It will
be used in the proof of Lemma \ref{le:reducible} (i).

 \bl \label{le:reducible=SUMirreducible,SEP,PPT}
Let $\r=\sum_i\r^{(i)}$ be an A or B-direct sum of the states
$\r^{(i)}$. Then $\r$ is separable [PPT] if and only if each
$\r^{(i)}$ is separable [PPT]. Consequently, $\r$ is a PPTES if and
only if each $\r^{(i)}$ is PPT and at least one of them is
entangled.
 \el

The set of normalized multipartite PPT states is a compact convex
set. We refer to its extreme points as {\em extreme states}. More
generally, for a non-normalized PPT state $\r$ we say that it is
{\em extreme} if the normalization of $\r$ is an extreme state. The
following definition generalizes that in \cite[Definition 4 ]{cd12}.

\bd \label{def:StronglyExtreme} The multipartite PPT state $\s$
is {\em strongly extreme} if there are no PPT states $\r\ne\s$
such that $\cR(\r)=\cR(\s)$.
 \ed
Obviously any pure product state is strongly extreme. By applying
the proof of \cite[Lemma 19]{cd12} to multipartite states, we obtain that any strongly extreme state $\r$ is extreme and that $\cR(\r)$ is a CES if $r(\r)>1$. It follows from Theorem
\ref{thm:3x3PPTstates} (iv) that any $3\times3$ PPTES of rank four is strongly extreme. In fact, we will show in Proposition
\ref{pp:stronglyextreme} that any PPTES of rank four is strongly
extreme.

\section{\label{sec:PPTrank23} Multipartite PPT states of rank two or three}

In this section we begin our investigation of multipartite PPT
states of small ranks. For PPT states of rank two or three, we will
show that they are separable (see Lemma \ref{le:PPTranktwo} and
Theorem \ref{thm:PPTrankthree}). We shall also determine their
lengths, generalizing the results obtained in \cite{cd12paper8} and
\cite{cd12PRA}.

Any PPT multipartite pure state is a product state and thus has
length one. Usually, multipartite states are assumed to have local
ranks larger than one. For the sake of completeness, we consider
also the states whose local ranks may be one. The following
observation is clear.
 \bl
 \label{le:rank1reduced}
Suppose $\r$ is an $n$-partite state with $r(\r_1)=1$ and so
$\r=\proj{a_1} \ox \s_{A_2\cdots A_n}$. Then
 \\ (i) $r(\r)=r(\s)$.
 \\ (ii) $\r$ is PPT if and only if $\s$ is PPT.
 \\ (iii) If $\r$ is separable, then so is $\s$ and
$L(\r)=L(\s)$.
 \el

Consequently, when dealing with the separablity of $\r$,
its rank, or the PPT property one can assume that all local
ranks are bigger than one.

We begin with states of rank two.
The following lemma generalizes \cite[Lemma 4]{kl01}.

 \bl
 \label{le:PPTranktwo}
Any multipartite PPT state of rank two is separable and has length two.
 \el
 \bpf
We use induction on $n$, the number of parties. The assertion is
trivial if $n=1$. Assume $n>1$ and let $\r$ be PPT state of rank
two. By Theorem \ref{thm:PPTMxNrank<M,N}, we must have
$r(\r_1)\le2$. If this rank is one, the assertion follows
immediately from the induction hypothesis. Hence, we may assume that $r(\r_1)=2$. By Proposition \ref{prop:PPTMxNrankN} (i) we
have
$\r=\proj{a}_{A_1}\ox\proj{\ph}_{A_2\cdots A_n}+
\proj{b}_{A_1}\ox\proj{\ps}_{A_2\cdots A_n}$, where
$\ket{a},\ket{b}$ are linearly independent. Since this is an
$A_1$-direct sum, the two summands on the right hand side must
be PPT states by Lemma \ref{le:rhoreducible=rhoPTreducible} and
so $\ket{\ph}$ and $\ket{\ps}$ are product vectors.
 \epf

The second assertion of this lemma has been obtained in \cite[Lemma
4 (iii)]{dfx07}. From this lemma we deduce another simple fact about
states of rank two. (This fact is also a corollary of 
\cite[Theorem 5]{cd11JPA}.)

 \bl
 \label{le:ranktwo=ent+sep}
If $\ket{\ph}\in\cH$ is a product vector and $\ket{\ps}\in\cH$
is entangled, then the state $\r=\proj{\ph}+\proj{\ps}$ is NPT.
 \el
 \bpf
Assume that $\r$ is PPT. By Lemma \ref{le:PPTranktwo} we have
$\r=\proj{\a}+\proj{\b}$, where $\ket{\a}=\ket{a_1,\ldots,a_n}$
and $\ket{\b}=\ket{b_1,\ldots,b_n}$ are product vectors.
Since $\ket{\ps}\in\cR(\r)$ is not a product vector, there are
at least two indexes $i$ such that $\ket{b_i}$
is not a scalar multiple of $\ket{a_i}$. It follows that
$\cR(\r)$ contains only two product vectors and so we have,
say, $\ket{\ph}=c\ket{\a}$. Thus
$\proj{\ps}=(1-|c|^2)\proj{\a}+\proj{\b}$. Since the right
hand side is positive semidefinite, we must have $|c|\le1$.
Hence, the right hand side is a separable state and we have
a contradiction. We conclude that $\r$ must be NPT.
 \epf

We need the following lemma about the bipartite case.
 \bl
 \label{le:length=Airreducible}
Let $\r=\sum^{l}_{i=0} \proj{a_i,b_i}$ be a bipartite A-irreducible
separable state of rank $r_1+1$ where $r_1:=r(\r_1)$. The equality
$L(\r)=r_1+1$ holds if (i) $r(\r^\G_1)=r_1+1$ or (ii)
$\ket{a_0}\propto\ket{a_1}$ and $\ket{b_0}$ and $\ket{b_1}$ are
linearly independent.
 \el
 \bpf
(i) Let $\s(t)=\r-t\proj{a_0,b_0}$ for real $t$. Since
$\r^\G_1=\sum^{l}_{i=0}\proj{a^*_i,b_i}$ and
$r(\r)=r(\r^\G_1)=r_1+1$, there is a $t_0\ge1$ such that $\s(t)\ge0$
and $\s(t)^\G_1\ge0$ for $t\le t_0$, and
$r(\s(t))=r(\s(t)^\G_1)=r_1+1$ for $t<t_0$ while at $t=t_0$ we have
$\min(r(\s(t_0)),r(\s(t_0)^\G_1))=r_1$. By Lemma
\ref{le:rhoreducible=rhoPTreducible} (iii), $\r^\G_1$ is also
$A$-irreducible. Thus, we may assume that $r(\s(t_0))=r_1$. The
equality $\r_1=\s(t_0)_1+t_0\|b_0\|^2\proj{a_0}$ and the fact that
$\r$ is $A$-irreducible imply that $r(\s(t_0)_1)=r_1$. By
Proposition \ref{prop:PPTMxNrankN}, $\s(t_0)$ is separable of length
$r_1$, and so $L(\r)=r_1+1$.

(ii) We may assume that $\ket{a_0}=\ket{a_1}$, and also that
$\{\ket{a_i}:1\le i\le r_1\}$ is a basis of $\cR(\r_1)$. As
$\ket{b_0}$ and $\ket{b_1}$ are linearly independent,
$\{\ket{a_i,b_i}:0\le i\le r_1\}$ is a basis of $\cR(\r)$. For
$j>r_1$ we have $\ket{a_j}=\sum_{i=1}^{r_1} \eta_i\ket{a_i}$ and
$\ket{a_j,b_j}=\sum_{i=0}^{r_1} \xi_i\ket{a_i,b_i}$, where
$\xi_i,\eta_i$ are some scalars. It follows that
$\xi_i\ket{b_i}=\eta_i\ket{b_j}$ for $1<i\le r_1$ and
$\xi_0\ket{b_0}+\xi_1\ket{b_1}=\eta_1\ket{b_j}$. If $\eta_i\ne0$ for
some $i>1$, then $\ket{b_j}\propto\ket{b_i}$. If $\eta_1\ne0$ then
$\ket{b_j}=\eta_1^{-1}(\xi_0\ket{b_0}+\xi_1\ket{b_1})$. Since
$\ket{a_j^*,b_j}=\sum_{i=1}^{r_1}\eta_i^*\ket{a_i^*,b_j}$, it follws
that $\ket{a_j^*,b_j}$ lies in the span of the vectors
$\ket{a_i^*,b_i}$, $i=0,\ldots,r_1$. Since this holds for all
$j>r_1$ and the vectors $\ket{a_i^*,b_i}$, $0\le i\le r_1$, are
linearly independent and belong to $\cR(\r^\G_1)$, we conclude that
$r(\r^\G_1)=r_1+1$. Hence, (ii) reduces to (i) and the proof is
completed.
 \epf

We often regard a multipartite state as a bipartite state by
considering the parties $A_2,\ldots,A_n$ as a single party.
The following lemma is useful in the analysis of such states.

 \bl
 \label{le:1(n-1)PPT}
Let $\r=\sum^l_{i=0} \proj{a_i}_{A_1} \ox
(\s^{(i)})_{A_2\cdots A_n}$ be a PPT state, $n\ge3$ and
$r_k:=r(\r_k)>1$ for each $k$.

(i) If $l=r_1-1$ and $r(\s^{(i)})\le2$ for each $i$, then each $\s^{(i)}$ is separable and  $L(\r)=r(\r)$.

(ii) If $r(\r)=r_1+1$, then $\r$ is separable. If we further
assume that $\r$ is $A_1$-irreducible, then $L(\r)=r_1+1$.
 \el
 \bpf
(i) Since $l=r_1-1$, the $\ket{a_i}$ are linearly independent.
Let $j\in\{0,\ldots,r_1-1\}$ be arbitrary and choose
$\ket{b_j}\in\cH_1$ such that $\braket{a_i}{b_j}=\d_{ij}$ for
all $i$. Since $\bra{b_j}\r\ket{b_j}=\s^{(j)}$, we conclude that $\s^{(j)}$ is PPT. By Lemma \ref{le:PPTranktwo}, $\s^{(j)}$ is separable and $L(\s^{(j)})=r(\s^{(j)})$. Since $j$
is arbitrary, $\r$ is separable. Since
$r(\r)=\sum^l_{i=0} r(\s^{(i)})=\sum^l_{i=0}
L(\s^{(i)})\ge L(\r)\ge r(\r)$, we have $L(\r)=r(\r)$.

(ii) We shall use induction on $r_1$. If $r_1=1$, the assertion is
true by Lemma \ref{le:PPTranktwo}. Now let $r_1>1$. Assume that $\r$
is $A_1$-reducible. Thus, we have $\r=\t\oplus_{A_1}\chi$. Since
$\r$ is biseparable for the partition $A_1:A_2\cdots A_n$, the same
is true for $\t$ and $\chi$. We have $r_1+1=r(\t)+r(\chi)$ and
$r_1=r(\t_1)+ r(\chi_1)$. Since $\r$ is PPT, the states $\t$ and
$\chi$ are also PPT. Theorem \ref{thm:PPTMxNrank<M,N} implies that,
say, $r(\t)=r(\t_1)$ and $r(\chi)=r(\chi_1)+1$. Then part (i) and
Proposition \ref{prop:PPTMxNrankN} imply that $\t$ is a sum of
$r(\t_1)$ pure bipartite product states. By part (i), $\t$ is
separable. So it remains to show $\chi$ is separable. This is true
when $r(\chi_1)=1$ by Lemma \ref{le:PPTranktwo}. So $r(\chi_1)>1$,
and the fact $r(\chi)=r(\chi_1)+1$ implies that there is at least
one $i>1$ such that $r_i>1$. If there is really only one such $i$,
then $\chi$ is separable. On the other hand if there are two
different $i,j>1$ such that $r_i,r_j>1$, then $\chi$ is separable by
induction hypothesis. So the assertion holds.


From now on we assume that $\r$ is $A_1$-irreducible.
Without any loss of generality, we may assume that
$\s^{(i)}=\proj{\psi_i}$ for each $i$. Note that we must have
$l\ge r_1$. We may also assume that the $\ket{a_i}$ with $i<r_1$
form a basis of $\cR(\r_1)$ and, by applying an ILO, we may assume that $\ket{a_i}=\ket{i}$ for $i<r_1$. We may assume that
the representation $\r=\sum_{i=0}^l \proj{a_i,\psi_i}$ is chosen
so that $l$ is minimal. In particular, no two $\ket{a_i,\psi_i}$
are parallel.

Suppose that $\r$ is entangled, and we will derive a
contradiction. Some $\ket{\psi_i}$ must be entangled, say
$\ket{\psi_0}$. For convenience, we can assume that
$r(\s^{(0)}_2)>1$. Since
 \bea \label{eq:LokalRang}
 r(\r_{12})
 \ge \sum_{i=0}^{r_1-1} r(\s^{(i)}_2)
 \ge 2 + (r_1-1) = r_1+1,
 \eea
we have $r(\r_{12})=r_1+1$ by Theorem \ref{thm:PPTMxNrank<M,N}. It follows that $r(\s^{(i)}_2)=1$ for $0<i<r_1$ and
$r(\s^{(0)}_2)=2$. Moreover, if $j\ge r_1$ and $\braket{i}{a_j}\ne0$ for some $0<i<r_1$ then we can replace $\s^{(i)}$ by
$\s^{(j)}$ in Eq. \eqref{eq:LokalRang}. Thus, we have
$r(\s^{(i)}_2)=1$ for all $i$ for which $\ket{a_i}$ is not
parallel to $\ket{0}$.

We claim that $l=r_1$. If two of the $\ket{a_i}$ are parallel, the
claim follows from Lemma \ref{le:length=Airreducible} (ii). We may
now assume that the $\ket{a_i}$ are pairwise non-parallel.
Consequently, $r(\s^{(i)}_2)=1$ for all $i>0$. Let us write
$\r=\proj{0,\ps_0}+\r'$ and let $\ket{\ps_i}=\ket{b_i,\ph_i}$ for
$i>0$. Since $r(\r_{12})=r_1+1$ and $\ket{0,\ps_0}$ is entangled
w.r.t. the bipartition $A_1A_2:A_3\cdots A_n$, \cite[Theorem
5]{cd11JPA} implies that $r(\r'_{12})=r_1+1$. Consequently, we may
assume that $\cR(\r)$ is spanned by the $\ket{a_i,b_i,\ph_i}$ with
$1\le i\le r_1+1$. We can now write
$\ket{0,\ps_0}=\sum^{r_1+1}_{i=1}x_i\ket{a_i,b_i,\ph_i}$, as well as
$\ket{a_{r_1}}=\sum_i c_i\ket{i}$ and $\ket{a_{r_1+1}}=\sum_i
d_i\ket{i}$. Since $\ket{a_i}=\ket{i}$ for $i<r_1$ and
$\ket{a_{r_1}}$ is not parallel to $\ket{0}$, we may assume that for
some $1\le k<r_1$ we have $c_i\ne0$ for $1\le i\le k$ and $c_i=0$
for $k<i<r_1$. So for $j=1,\ldots,k$, we have
 \bea
 \ket{\ps_{0}}
 &=&
 x_{r_1}c_0\ket{b_{r_1},\ph_{r_1}}+x_{r_1+1}d_0\ket{b_{r_1+1},\ph_{r_1+1}},
 \\
 -x_j\ket{b_j,\ph_j}
 &=&
 x_{r_1}c_j\ket{b_{r_1},\ph_{r_1}}+x_{r_1+1}d_j\ket{b_{r_1+1},\ph_{r_1+1}}.
 \eea
Since $\ket{\ps_0}$ is entangled w.r.t. the partition
$A_2:A_3\cdots A_n$, we have $x_{r_1}x_{r_1+1}c_0d_0\ne0$. As
$c_j\ne0$, we must have $x_j\ne0$, $d_j=0$ and
$\ket{b_j,\ph_j}\propto\ket{b_{r_1},\ph_{r_1}}$. Thus
 \bea
 \r
 &=&
 \proj{0,\ps_0}+\sum^k_{j=1}\proj{a_j,b_j,\ph_j}+\proj{a_{r_1},b_{r_1},\ph_{r_1}}+\r''
 \notag\\
 &=&
 \proj{0,\ps_0}+(y\proj{0}+\chi)\proj{b_{r_1},\ph_{r_1}}+\r'',
 \eea
with $y\ne0$, $\chi$ positive semidefinite, and $\r''$ biseparable
for the partition $A_1:A_2\cdots A_n$. Now Lemma
\ref{le:length=Airreducible} (ii) shows that $l$ must be equal to
$r_1$. Thus, our claim is proved.

Since $l=r_1$ and $\r$ is $A_1$-irreducible, we obtain
$r(\s^{(i)}_2)=1$ for $i>0$. Lemma \ref{le:ranktwo=ent+sep} implies
that the state $\bra{0}_{A_1}\r\ket{0}_{A_1}$ is NPT with respect to
the partition $A_2:A_3\cdots A_m$. Hence, we have a contradiction.
Thus, $\r$ must be separable.

Finally we prove the second assertion of (ii). Let
$\ket{\ps_i}=\ket{a_{i,2},\ldots,a_{i,n}}$ for all $i$. By Theorem
\ref{thm:PPTMxNrank<M,N} and Proposition \ref{prop:PPTMxNrankN},
$r(\r_{12})=r_1$ or $r_1+1$. In the former case, $\cR(\r_{12})$ is
spanned by the $\ket{a_i,a_{i,2}}$, $i < r_1$. Since any
$\ket{a_i,a_{i,2}}\in\cR(\r_{12})$ and $r_2>1$, we see that $\r$ is
$A_1$-reducible and we have a contradiction. Thus
$r(\r_{12})=r_1+1$, and we may assume that $\cR(\r)$ is spanned by
the $\ket{a_i,\ps_i}$, $i \le r_1$. Using an ILO we may assume
$\ket{a_{r_1}}=\sum^{s-1}_{i=0}\ket{i}$, $s\le r_1$. We divide the
set of integers $0,1,\cdots, r_1$ into $k\ge2$ disjoint subsets
$S_1,\cdots,S_k$. Any two vectors $\ket{a_{i,3},\ldots,a_{i,n}}$ and
$\ket{a_{j,3},\ldots,a_{j,n}}$ are linearly independent if and only
if $i,j$ are from different subsets $S_i,S_j$. Since any
$\ket{a_i,\ps_i}\in\cR(\r)$, Proposition \ref{prop:PPTMxNrankN}
implies that for $j>r_1$, we have
$\ket{a_{j,3},\ldots,a_{j,n}}\propto\ket{a_{i,3},\ldots,a_{i,n}}$
for some $i\le r_1$. The state can be written as
 \bea
 \r = \sum^k_{i=1} \bigg( \r^{(i)} \ox \proj{a_{i,3},\cdots,a_{i,n}}
 \bigg),
 \eea
where $\r^{(i)}$ are bipartite separable states and $\cR(\r^{(i)})$
is spanned by $\ket{a_i,a_{i2}},i\in S_i$. So $\sum_i
r(\r^{(i)})=r(\r_{12})=r(\r)$.

Since $\r$ is $A_1$-irreducible and $k\ge2$, no set $S_i$ consists
of $0,\cdots,s-1,r_1$. So $r(\r^{(i)})=r(\r^{(i)}_1)$ for any $i$.
We have $L(\r^{(i)})=r(\r^{(i)})$ by Proposition
\ref{prop:PPTMxNrankN}. Therefore $r(\r) \le L(\r)\le \sum_i
L(\r^{(i)})=r(\r)$. So the proof for the second assertion of (ii) is
completed.
 \epf

We shall see later (see Lemma \ref{le:reducible}) that the
inequalities $r(\s^{(i)})\le2$ in part (i) of the above lemma can be
replaced with $r(\s^{(i)})\le3$. We also point out that the length
of a reducible state may be bigger than its rank. An example is the
tripartite separable state
$\r=(\proj{000}+\proj{111})\op_{A_1}(\s\ox\proj{0})$, where $\s$ is
a two-qubit separable state of birank $(3,4)$ (see \cite[Table
1]{cd12PRA}). So $r(\r)=r_1+1=5<L(\r)=r(\r^\G_1)=6$. Hence the
second assertion of (ii) fails for reducible states.

Next, we consider states of rank three. We shall prove that, for any number of parties, any PPT state of rank three is separable.

 \bt
 \label{thm:PPTrankthree}
Let $\r$ be an $n$-partite PPT state of rank three, $n>1$, and
let $r_k:=r(\r_k)>1$ for each $k$. Then

(i) $\r$ is separable;

(ii) $L(\r)\in\{3,4\}$ and if $L(\r)=4$ then $n=2$ and
$r_1=r_2=2$.
 \et
 \bpf
By Theorem \ref{thm:PPTMxNrank<M,N}, we have $r_i\le3$ and so
$r_i\in\{2,3\}$ for each $i$.

(i) Proposition \ref{prop:PPTMxNrankN} (i) implies that
$\r=\sum_i \proj{a_i}_{A_1}\ox (\s^{(i)})_{A_2\cdots A_n}$. Thus
(i) holds if $n=2$, so let $n>2$. If $r_1=3$, (i) follows from Proposition \ref{prop:PPTMxNrankN} (i) and Lemma
\ref{le:1(n-1)PPT} (i). If $r_1=2$, (i) follows from Lemma
\ref{le:1(n-1)PPT} (ii). Hence, (i) has been proved.

(ii) Recall that the length of a separable state is always greater than or equal to its rank. Thus we have $L(\r)\ge3$. Suppose $\r$ is reducible, say $\r=\t\op_{A_1}\c$. Since
$r(\t),r(\c)\le2$, we have $3\le L(\r) \le L(\t)+ L(\c)=3$ by
Lemma \ref{le:PPTranktwo}. If $r_1=3$ then $L(\r)=3$ by
Proposition \ref{prop:PPTMxNrankN} (i). So we can assume $\r$ is
irreducible and $r_i=2$ for all $i$. If $n\ge3$ then $L(\r)=3$ by Lemma \ref{le:1(n-1)PPT} (ii). Finally, if $n=2$ then (ii) follows from \cite[Table 1]{cd12PRA}.
 \epf

Part (i) may be restated as follows: any $n$-partite PPTES must have rank at least four.

While bipartite separable states of rank two always have length two (see Lemma \ref{le:PPTranktwo}), the ones of rank three may have length four (see \cite[Table 1]{cd12PRA}).

\section{\label{sec:PPTrank4} Multipartite PPT states of rank four}

We now begin the study of states of rank four. The main result
is Theorem \ref{thm:PPTrankfour}, where we show that there exist
only two types of multipartite PPTES of rank four. They are
either $2\times2\times2$ or $3\times3$ states.

Let us begin with the reducible PPT states.
 \bl
 \label{le:reducible}
(i) Any reducible $2\times2\times2$ or $2\times2\times3$ PPT
state is separable.

(ii) Any reducible $n$-partite PPT state of rank four is separable of length at most five. The bound five is sharp.

(iii) The first assertion of Lemma \ref{le:1(n-1)PPT} (i) remains
valid when we replace ``$r(\s^{(i)})\le2$'' with
``$r(\s^{(i)})\le3$''.
 \el
 \bpf
(i) Suppose $\r$ is a reducible $2\times2\times2$ or
$2\times2\times3$ PPT state, say $\r=\a\op_{A_1}\b$. Since $\r$ is
PPT, so are $\a$ and $\b$ (see Lemma
\ref{le:reducible=SUMirreducible,SEP,PPT}). Hence, they are
separable by the Peres-Horodecki criterion. The case where $\r$ is a $2\times2\times3$ PPT state and $\r=\a\op_{A_3}\b$ can be treated similarly.

(ii) Suppose $\r$ is a reducible PPT state of rank four, say
$\r=\a\op_{A_1}\b$. Then both $\a$ and $\b$ are PPT and
$r(\a)+r(\b)=4$. By Lemma \ref{le:PPTranktwo} and Theorem
\ref{thm:PPTrankthree}, $\a$ and $\b$ are separable and
$L(\r)\le L(\a)+L(\b)\le\max(2+2,4+1)=5$. The bound five is
reached by the tripartite separable state $\r=\proj{000}\op_{A_1}(\s\ox\proj{1})$, where $\s$ is a two-qubit separable state of birank $(3,4)$ (see \cite[Table 1]{cd12PRA}).

(iii) This follows from the proof of Lemma \ref{le:1(n-1)PPT} (i) by using Theorem \ref{thm:PPTrankthree} (i).
 \epf

Let us point out that reducible $2\times3\times3$ PPT states
as well as reducible bipartite PPT states of rank five may be entangled. As examples, we can take respectively the states
$(\proj{0}\ox\s)\op_{A_1}\proj{100}$ and $\s\op_{A_1}\proj{30}$,
where $\s$ is a bipartite $3\times3$ PPTES of rank four.

In analogy with Lemma \ref{le:1(n-1)PPT}, we can consider
tripartite PPT states
$\r=\sum_i \proj{\ps_i}_{A_1A_2} \ox \proj{c_i}_{A_3}$ with
$r(\r)=r(\r_{1,2})+1$. However such states may be entangled.
As an example we can take the state
$\r=\s\ox\proj{0}+\e\proj{00}\ox(\proj{0}+\proj{1})$,
where $\s$ is a $3\times3$ PPTES of rank four and $\e>0$
is small. Indeed, such $\r$ is a $3\times3\times2$ PPTES with
$r(\r)=6$ and $r(\r_{1,2})=5$.

We can now obtain sharp upper bounds for the lengths of separable states of rank four.
 \bl
 \label{le:sepRANK4length}
 Let $\r$ be an $n$-partite separable state of rank four. For the sake of simplicity assume that $n>1$ and that the ranks
$r_i:=r(\r_i)$ satisfy the inequalities
$1<r_1\le r_2\le\cdots\le r_n$.

 (i) If $n=2$, then $L(\r)\le6$ and equality holds only if
 $\r$ is a $2\times3$ state.

 (ii) If $n>2$ and $r_n>2$, then $L(\r)\le5$ and equality holds only if
 $\r$ is reducible.

 (iii) If $n>2$ and $r_n=2$, then $L(\r)\le6$ and equality holds only if $\r$ is the sum of pure product states whose generating product vectors are in general position.
 \el
 \bpf
Since $\r$ is separable, we have $r_n\le4$. If $r_n=4$ then
$L(\r)=4$ by Proposition \ref{prop:PPTMxNrankN} (i) and Lemma
\ref{le:1(n-1)PPT} (i). Thus, we may assume that $r_n\le3$.

(i) If $r_1=3$ then $L(\r)\le5$ by \cite[Lemma 16]{cd12PRA}. If
$r_2=2$ then $L(\r)\le4$ by \cite[Table I]{cd12PRA}. If $r_1=2$ and
$r_2=3$ then $L(\r)\le6$ by \cite[Table II]{cd12PRA}, where an
example with $L(\r)=6$ was constructed.

(ii) As we assumed that $r_n\le3$, we must have $r_n=3$. For
irreducible $\r$, after switching the parties $A_1$ and $A_n$, Lemma
\ref{le:1(n-1)PPT} (ii) implies that $L(\r)=4$. On the other hand
for reducible $\r$, the assertion follows from Lemma
\ref{le:reducible} (ii).

(iii) We may assume now that each $\cH_i$ has dimension two. If $\r$ is reducible then $L(\r)\le5$ by Lemma
\ref{le:reducible} (ii). So, we assume that $\r$ is irreducible.
Let $m$ be an integer such that $1<m<n$. For
$1\le i_1<\cdots<i_m\le n$ we shall denote by
$r_{i_1,\ldots,i_m}$ the rank of the reduced state
$\r_{i_1,\ldots,i_m}$. We can write
$\r=\sum^l_{i=1}\proj{\psi_i}$, where $l\ge4$ and the
$\ket{\psi_i}=\ket{a_{i,1},\ldots,a_{i,n}}$ are product vectors.
Note that $r_{i_1,\ldots,i_m}\in\{2,3,4\}$ and so we shall
consider three cases.

Case 1:  $r_{1,\ldots,m}=2$ for some choice of indexes
$i_1<\cdots<i_m$. We may assume that $i_k=k$ for $k=1,\ldots,m$.
Then $\cR(\r_{1,\ldots,m})$ is spanned by two of the
$\ket{\psi_i}$, say those for $i=1,2$. Since each $r_i=2$, the
vectors $\ket{a_{1,j}}$ and  $\ket{a_{2,j}}$ must be linearly
independent for each $j\in\{1,\ldots,m\}$.
Consequently, there are no other product vectors in
$\cR(\r_{1,\ldots,m})$. It follows that $\r$ is $A_1$-reducible and we have a contradiction.

Case 2:  $r_{1,\ldots,m}=4$ for some choice of indexes
$i_1<\cdots<i_m$. We may assume that $i_k=k$ for $k=1,\ldots,m$
and that $m=n-1$.
Let $\ket{\psi'_i}=\ket{a_{i,1},\ldots,a_{i,n-1}}$ and note that
$\cR(\r_{1,\ldots,m})$ is spanned by four $\ket{\psi'_i}$,
say those for $i=1,2,3,4$. It follows that the $\ket{\psi_i}$,
$i=1,2,3,4$, span $\cR(\r)$. Consequently, at least two of the
$\ket{a_{i,n}}$, $i=1,2,3,4$, are linearly independent. Let $k$
be the maximum number of pairwise non-parallel vectors among
$\ket{a_{i,n}}$, $i=1,2,3,4$. As $r_n=2$ we have $k>1$. Since
$\r$ is irreducible, we must have $k>2$. It follows that
$\r=\sum^k_{s=1}\r^{(s)}\ox\proj{a_{s,n}}$, where each
$\r^{(s)}$ is an $(n-1)$-partite separable state and
$r(\r^{(1)})+\cdots+r(\r^{(k)})=4$. As $k>2$, we have
$r(\r^{(s)})\le2$ for all $s$. By Lemma \ref{le:PPTranktwo}, we
have $L(\r)\le\sum^k_{s=1} L(\r^{(s)})\le4$.

Case 3: $r_{i_1,\ldots,i_m}=3$ for any sequence
$i_1<\cdots<i_m$. We may assume that the $\ket{\psi_i}$,
$i=1,2,3,4$, span $\cR(\r)$. For each $j\in\{1,\ldots,n\}$
denote by $k_j$ the maximum size of a subset
$X_j\sue\{1,2,3,4\}$ such that all $\ket{a_{i,j}}$ with
$i\in X_j$ are parallel to each other. Let
$k=\max(k_1,\ldots,k_n)$ and note that $k<4$. Without any loss
of generality we may assume that $k_1=k$ and that
$\ket{a_{1,1}}=\cdots=\ket{a_{k,1}}$. For any $i$ let
$\ket{\ps''_i}=\ket{a_{i,2},\ldots,a_{i,n}}$. We have three
subcases.

First, let $k=3$. Since $r_{2,3,\ldots,n}=3$, the
vectors $\ket{\ps''_i}$, $i=1,2,3$, span
$\cR(\r_{2,3,\ldots,n})$. If $\ket{a_{i,1}}$ and $\ket{a_{1,1}}$
span $\cH_1$, then $\ket{\ps''_i}\propto\ket{\ps''_4}$.
Thus, we have $\r=\proj{a_1}\ox\r''+\r'\ox\proj{\ps''_4}$ for
some $(n-1)$-partite separable state $\r''$ of rank three and
some $\r'\ge0$. Since $\r'=p\proj{a_1}+q\proj{a'}$ where
$p\ge0$, $q>0$ and $\ket{a'}\ne0$, $\r$ is $A_1$-reducible,
and so we have a contradiction.

Second, let $k=2$. We may assume that
 \bea \label{eq:KomplRo}
 \ket{\ps_1} &=& \ket{0,\ldots,0,0,\ldots,0},
 \\
 \ket{\ps_2} &=& \ket{0,\ldots,0,1,\ldots,1},
 \\
 \ket{\ps_3} &=& \ket{1,\ldots,1,b_{s+1},\ldots,b_n},
 \\
 \ket{\ps_4} &=& \ket{b_1,\ldots,b_s,b_{s+1}',\ldots,b_n'}.
 \eea
As $k=2$, $\ket{0}$ and $\ket{b_i}$ must be linearly independent
for $i\le s$. Since $r_{1j}=3$ for $j>s$, it follows that
$\ket{b'_{s+1},\ldots,b'_n}\propto\ket{b_{s+1},\ldots,b_n}$. If
$s=1$ then, for suitably chosen $\ket{a}\in\cH_1$, we have
 \bea
\proj{0,\ps''_3}+\proj{a,\ps''_3}=\proj{\ps_3}+\proj{\ps_4}.
 \eea
Thus, we can replace $\ket{\ps_3}$ and $\ket{\ps_4}$ with
$\ket{0,\ps''_3}$ and $\ket{a,\ps''_3}$ and so we obtain the first
subcase $(k=3)$. Similarly, if $s=n-1$ we can reduce the problem to
the first subcase. Now let $1<s<n-1$. Since $r_{1,\ldots,s}=3$, at
least one of the vectors $\ket{b_j}$ with $j\le s$ is not parallel
to $\ket{1}$. This condition implies $r_{1,\cdots,s+1}=4$ and it is
a contradiction.


Third, let $k=1$. So the product vectors $\ket{\ps_i}$ are in
general position. Note that in all instances covered so far we had
$L(\r)\le5$. By applying an ILO and by using the fact that
$r_{i_1,\ldots,i_m}=3$ for $1<m<n$, we may assume that
 \bea \label{eq:VektoriPsi}
 \ket{\ps_1}=\lambda\ket{0,\ldots,0},\quad
 \ket{\ps_2}=\mu\ket{1,\ldots,1},\quad
 \ket{\ps_3}=\ket{e,\ldots,e},
 \eea
where $\ket{e}=\ket{0}+\ket{1}$ and $\lambda\mu\ne0$. (We have here
identified all the spaces $\cH_j$ by using their bases
$\{\ket{0},\ket{1}\}$.) For $j\ne1$ we have $r_{1j}=3$, and so
$\ket{a_{i1},a_{ij}}$ is a linear combination of $\ket{00}$,
$\ket{11}$ and $\ket{ee}$. An easy computation shows that
$\ket{a_{ij}}\propto\ket{a_{i1}}$ must hold. Consequently, $\cR(\r)$
is spanned by symmetric product vectors. Since $r_{2,\ldots,n}=3$
and $r(\r)=4$, Proposition \ref{prop:PPTMxNrankN} (i) implies that
$4\le r(\r^{\G_1})\le6$. Let $l$ be the length of $\r$ when viewed
as a bipartite separable state for the partition $A_1:A_2\cdots
A_n$. By \cite[Proposition 12]{cd12PRA}, we have
$l=r(\r^{\G_1})\le6$. Thus, $\r=\sum^l_{i=1} \proj{v_i}_{A_1}\ox
\proj{\chi_i}_{A_2\cdots A_n}$. Since each $\ket{v_i,\chi_i}$ is a
linear combination of symmetric product vectors, it follows that
$\ket{v_i,\chi_i}$ is also a symmetric product vector. Hence,
$L(\r)\le l\le6$. The equality $L(\r)=6$ may hold. For example, we
have $L(\r)=r(\r^{\G_1})=6$ for the state
 \bea
 \r = \proj{000} + \proj{111}
 + \sum^4_{j=1} \proj{e_j,e_j,e_j},
 \eea
where $\ket{e_1}=\ket{0}+\ket{1}$, $\ket{e_2}=\ket{0}-\ket{1}$,
$\ket{e_3}=\ket{0}+e^{\frac13\p i}\ket{1}$ and
$\ket{e_4}=\ket{0}+e^{\frac23\p i}\ket{1}$.
This completes the proof.
 \epf

 \bl
 \label{le:3x3x3PPT}
Any $3\times3\times3$ PPT state of rank four is separable.
 \el
 \bpf
Let $\r$ be such a state and set $r_{ij}=r(\r_{ij})$. Suppose that
$r_{23}=r_{13}=r_{12}=3$. Then by Proposition \ref{prop:PPTMxNrankN}
(i) we have $\r_{12}=\sum^3_{i=1}\proj{a_i,b_i}$ and
$\r_{23}=\sum^3_{i=1}\proj{b_i',c_i}$. As $\r$ is a
$3\times3\times3$ state, each of the triples $\{\ket{a_i}\}$,
$\{\ket{b_i}\}$, $\{\ket{b_i'}\}$, $\{\ket{c_i}\}$ spans a
3-dimensional subspace. Consequently, $\r$ can be written (see
\cite[p7]{cw04}) as
 \bea
 \r = \sum_i
 \bigg( \sum^3_{j=1}\ket{a_j,b_j,c'_{i,j}} \bigg)
 \bigg( \sum^3_{j=1}\bra{a_j,b_j,c'_{i,j}} \bigg).
 \eea
Note that $\cR(\r_{23})$ contains only three product vectors, namely
the $\ket{b'_i,c_i}$. By tracing out system $A_1$, we deduce that
all $\ket{b_j,c'_{i,j}}\in\cR(\r_{23})$. For any $j$ there is at
least one $i$ such that $\ket{c'_{i,j}}\ne0$. By permuting the
$\ket{b'_j,c_j}$, we may assume that $\ket{b_j}=\ket{b'_j}$ for each
$j$, and that each $\ket{b_j,c'_{i,j}}$ is a scalar multiple of
$\ket{b_j,c_j}$. Hence, we have
$\r=\sum^3_{j=1}p_j\proj{a_j,b_j,c_j}$, $p_j>0$. As $\r$ has rank
four, we have a contradiction. Thus some $r_{ij}=4$, say $r_{23}=4$.
Then it follows from Proposition \ref{prop:PPTMxNrankN} (i) that
$\r$ satisfies the conditions of Lemma \ref{le:1(n-1)PPT}. Hence,
$\r$ is separable by part (ii) of that lemma.
 \epf

 \bl
 \label{le:2x3x3,2x3x2PPT}
Any $2\times3\times2$ or $2\times3\times3$ PPT state of rank
four is separable.
 \el
 \bpf
Let $\r$ be a $2\times3\times s$ PPT state of rank four with
$s\in\{2,3\}$ and set $r_{ij}=r(\r_{ij})$.
By Theorem \ref{thm:PPTMxNrank<M,N} we have
$r_{12},r_{23}\in\{3,4\}$ and $r_{13}\in\{2,3,4\}$.
Let us view $\r$ as a bipartite state for the partition
$A_1:A_2A_3$. Then the last assertion of
Proposition \ref{prop:PPTMxNrankN} (i) implies that
 \bea \label{eq:Split1:23}
 \r = \sum^m_{i=1} \proj{a_i}_{A_1} \ox \r^{(i)}_{A_2A_3},
 \eea
where the $\ket{a_i}$ are pairwise linearly independent.
By Lemma \ref{le:reducible} (ii), $\r$ is separable if $m=2$.
Thus, we assume that $m>2$.

Let us first consider the case $r_{12}=3$.
Assume that some $\r^{(i)}$, say $\r^{(1)}$, is entangled.
Then $\cR(\r_{12})$ has a basis
$\{\ket{a_1,x},\ket{a_1,y},\ket{a_2,z}\}$.
Consequently, for any product vector $\ket{u,v}\in\cR(\r_{12})$,
$\ket{u}$ must be a scalar multiple of $\ket{a_1}$ or
$\ket{a_2}$. It follows from Eq. \eqref{eq:Split1:23} that
$\ket{a_3,w}\in\cR(\r_{12})$ for some $\ket{w}\in\cH_2$,
and so we have a contradiction.
We conclude that all $\r^{(i)}$ must be separable and so $\r$ is
separable.

From now on we consider the remaining case $r_{12}=4$. If $s=3$
then, by Proposition \ref{prop:PPTMxNrankN} (i), $\r$ is
biseparable for the partition $A_1A_2:A_3$ and it is separable by
Lemma \ref{le:1(n-1)PPT}. Thus, we are done with the case $s=3$.
We continue with the case $s=2$. In view of Lemma
\ref{le:reducible} (ii), we may assume that $\r$ is irreducible.
If $r_{13}=4$ then Proposition \ref{prop:PPTMxNrankN} (i), and
Lemma \ref{le:1(n-1)PPT} imply that $\r$ is separable. Thus, we
assume from now on that $r_{13}<4$. By Proposition
\ref{prop:PPTMxNrankN} (i) we have
 \bea
 \r=\sum^3_{i=0}\proj{\ps_i}_{A_1A_2}\ox(\proj{c_i})_{A_3}.
 \eea

Assume that some $\ket{\ps_i}$, say $\ket{\ps_0}$, is entangled.
Then $\cR(\r_{23})$ contains three linearly independent vectors
$\ket{x,c_0}$, $\ket{y,c_0}$ and $\ket{b,c_1}$, where we assume
(as we may) that $\ket{c_1}$ is not parallel to $\ket{c_0}$. If
$r_{23}=3$, then any product vector in $\cR(\r_{23})$ is
parallel to $\ket{b,c_1}$ or equal to $\ket{z,c_0}$ for some
$\ket{z}\in\cH_2$. This contradicts the irreducibility of $\r$.
Hence, we must have $r_{23}=4$. If $\ket{c_0}$ is not parallel to
any $\ket{c_i}$, $i>0$, then $r_{13}<4$ implies that for $i>0$ we have $\ket{\ps_i}=\ket{a,b_i}$. As $r_{12}$ is PPT,
\cite[Theorem 5]{cd11JPA} gives a contradiction.
Consequently, we may assume that $\ket{c_3}=\ket{c_0}$. Since
$\r$ is irreducible, the $\ket{c_i}$ with $i<3$ are pairwise linearly independent. As $r_{13}<4$, we must have
$\ket{\ps_i}=\ket{a,b_i}$ for $i=1,2$. Let
$\s=\proj{\ps_0}+\proj{\ps_3}$. As $\ket{\ps_0}$ is entangled,
we have $r(\s_2)\in\{2,3\}$. If $r(\s_2)=3$ then $r_{23}=4$
implies that $\ket{b_1}\propto\ket{b_2}$ contradicting the fact
that $\r$ is irreducible. Hence, we must have $r(\s_2)=2$. Then
we may assume that $\ket{\ps_3}$ is a product vector, and we
introduce the separable state $\r'=\r-\proj{\ps_0,c_0}$.
Since $r_{23}=4$ and $r(\r'_{23})=3$, \cite[Theorem 5]{cd11JPA}
gives a contradiction.

Hence, we have shown that all $\ket{\ps_i}$
must be product vectors and so $\r$ is separable.
 \epf

 \bl
 \label{le:m(n-m)PPT}
Suppose $\r$ is an $n$-partite PPT state of rank four, $n>2$,
all ranks $r_i:=r(\r_i)>1$, and $\sum^n_{i=1}r_i>6$. Then $\r$ is
separable.
 \el
 \bpf
Theorem \ref{thm:PPTMxNrank<M,N} implies that $r_i\le4$. If some
$r_i=4$, then $\r$ is separable by Proposition
\ref{prop:PPTMxNrankN} (i) and Lemma \ref{le:1(n-1)PPT} (i). If some
$r_i=3$, we can combine the system
$A_2,\cdots,A_{i-1},A_{i+1},\cdots,A_n$ into a single party $B$.
Then Proposition \ref{prop:PPTMxNrankN} (i) implies that $4\ge
r(\r_B)\ge2$. We claim that the tripartite state $\r_{A_1A_iB}$ is
separable. The case $r(\r_B)=4$ has been discussed before. For the
case $3\ge r(\r_B)\ge2$, we use the fact $2\le r_1\le3$ and apply
Lemma \ref{le:3x3x3PPT} and \ref{le:2x3x3,2x3x2PPT}. So the claim
holds. By replacing $A_i$ by $A_1$ in Lemma \ref{le:1(n-1)PPT} (ii),
$\r$ is separable. So it is sufficient to consider $r_i=2$ for all
$i$. Then the condition $\sum^n_{i=1}r_i>6$ implies that $n>3$.
Furthermore we may assume $\r$ is irreducible by Lemma
\ref{le:reducible}. For $1\le i_1<\cdots<i_m\le n$ we shall denote
by $r_{i_1,\ldots,i_m}$ the rank of the reduced state
$\r_{i_1,\ldots,i_m}$. We have $r_{i_1,\ldots,i_m}\le4$ by Theorem
\ref{thm:PPTMxNrank<M,N}. Since $\r$ is PPT, we deduce that all
$r_{ij}\in\{2,3,4\}$.

We show that the cases that some $r_{ij}=2$ or $4$ can be reduced to
the case that some $r_{ij}=3$. Suppose some $r_{ij}=4$, say
$r_{12}=4$. By Proposition \ref{prop:PPTMxNrankN} (i) we have
$\r=\sum^4_{i=1}\proj{\a_i}_{A_1A_2}\ox\proj{\b_i}_{A_3\cdots A_n}$.
By combining $A_1$ and $A_2$ into a single party and by applying
Lemma \ref{le:1(n-1)PPT} (i), we conclude that each $\ket{\b_i}$ is
a product vector. Since $\r$ is irreducible, we may assume
$r_{34}=3$ or $4$. If $r_{34}=4$, then the assertion will follow
from $r_{2,\cdots,n}\le4$. So it is sufficient to consider
$r_{34}=3$. Next, suppose some $r_{ij}=2$, say $r_{12}=2$.
Proposition \ref{prop:PPTMxNrankN} (i) implies that $\r=\sum_i
\proj{\ps_i'}_{A_1A_2}\ox\r^{(i)}_{A_3:\cdots:A_n}$. Since $\r$ is
an irreducible multiqubit PPT state, there is at least one entangled
state $\ket{\ps_i'}$. So $r_{23}=3$ or $4$. By these arguments we
have proved that it is sufficient to consider $r_{ij}=3$, say
$r_{23}=3$.

Let us view $\r$ as a $(n-1)$-partite PPT state $\s$ for the
partition $A_2A_3:A_1:A_4:\cdots:A_n$. Since $n>3$, we may replace
$r_{23}$ by $r_i$ in the first paragraph of the proof. So $\s$ is
separable, we have
$\r=\s=\sum^k_{i=1}\proj{\ps_i}_{A_2A_3}\ox\proj{\ph_i}_{A_1A_4\cdots
A_n}$, where $\ket{\ph_i}$ are product vectors.

We claim that $k=4$. Since $\r$ is irreducible, the state $\s$ is
reducible if and only if it is $A_2A_3$-reducible. We have
$\s=\a\op_{A_2A_3}\b$ where $\a,\b$ are $(n-1)$-partite separable
states w.r.t. to the partition $A_2A_3:A_1:A_4:\cdots:A_n$. There
are only two cases, i.e., $(r(\a),r(\b))=(3,1)$ or $(2,2)$. In the
former case, the fact $\r$ is an irreducible multiqubit PPT state
implies that $r(\a_{23})=r(\a_i)=2$, $i=1,4,\cdots,n$. Recall that
$n\ge4$, Theorem \ref{thm:PPTrankthree} implies $L(\s)\le
L(\a)+L(\b)=3+1=4$. In the latter case, Lemma \ref{le:PPTranktwo}
implies that $L(\s)\le L(\a)+L(\b)=2+2=4$. Since $L(\s)\ge r(\s)=4$,
we have $k=4$ for the reducible state $\s$. On the other hand if
$\s$ is irreducible, the condition $r_{23}=3$ and Lemma
\ref{le:1(n-1)PPT} (ii) imply $k=4$ too. So the claim has been
proved.

Assume some state $\ket{\ps_i}$, say $\ket{\ps_1}$ is entangled. The
fact $\r$ is PPT indicates that $r_{3\cdots n}=3$ and the states
$\ket{\ph_i}$ are pairwise linearly independent. Suppose
$\ket{\ph_i}$, $i=1,2,3$ form a basis of $\cR(\r_{3\cdots n})$, and
we choose $\ket{\ph}\perp\lin\{\ket{\ph_2},\ket{\ph_3}\}$. So the
bipartite state $\bra{\ph}\r\ket{\ph}=a\proj{\ps_1}+b\proj{\ps_4}$,
$a>0,b\ge0$ is PPT. Since $\ket{\ps_1}$ is entangled, Lemma
\ref{le:ranktwo=ent+sep} implies that $b>0$ and $\ket{\ps_4}$ is
entangled. Since the states $\ket{\ph_i}$ are pairwise linearly
independent, we have $r_{2,\cdots,n}>4$ and it is a contradiction
with Theorem \ref{thm:PPTMxNrank<M,N}. Thus all $\ket{\ps_i}$ must
be separable and so is $\r$.
 \epf

The following is an easy consequence:
 \bcr
 \label{cr:4partite}
An $n$-partite state $\r$ of rank four, with $n\ge4$ and all
$r(\r_i)>1$, is separable if and only if it is PPT.
 \ecr

Now we can present the main result of this section.

 \bt
 \label{thm:PPTrankfour}
Let $\r$ be an $n$-partite PPTES of rank four with all
$r_i:=r(\r_i)>1$. Then either $n=2$ and $r_1=r_2=3$ or
$n=3$ and $r_1=r_2=r_3=2$. Conversely, in these two cases
such PPTES exist.
 \et
 \bpf
If $n=2$ then Proposition \ref{prop:PPTMxNrankN} implies that $\r$
is a $3\times3$ state. If $n>2$ then Lemma \ref{le:m(n-m)PPT}
implies that $\r$ is a $2\times2\times2$ state. For the existence
assertion see \cite{DiV03}.
 \epf

To conclude this section, we present a lemma and a conjecture
beyond the scope of this section.

 \bl
 \label{le:rank<=r}
Suppose all $d_1 \times \cdots \times d_n$ PPT states of rank $r$
are separable. Then all $d_1 \times \cdots \times d_n$ PPT states of rank $\le r$ are separable.
 \el
 \bpf
Suppose $\r$ is a $d_1\times\cdots\times d_n$ PPT state of rank
less than $r$. We can choose a separable state $\s$ such that
$\r+\s$ has rank $r$. Then for any $t>0$ the state $\r+t\s$ is PPT of rank $r$, and so it is separable by the hypothesis. As
the set of separable states is closed, it follows that $\r$ is separable.
 \epf

 \bcj
Suppose all $d_1\times\cdots\times d_n$ PPT states of rank $r$,
with each $d_i>1$ and $n\ge3$, are separable. Then

(i) all $(d_1+1)\times d_2\times\cdots\times d_n$ PPT states of
rank $r$ are separable;

(ii) all $d_1\times\cdots\times d_n\times d_{n+1}$ PPT states
of rank $r$ are separable.
 \ecj
Note that both assertions are false for $n=2$. Indeed, all
$2\times3$ and $2\times2$ PPT states are separable but there exist $3\times3$ and $2\times2\times2$ PPTES of rank four.

\section{\label{sec:3qubit} Three-qubit PPTES of rank four}

In this section we study the properties of three-qubit PPTES of rank
four, see Propositions \ref{pp:PPTESrankAB=4} and
\ref{pp:3qubitPPTES,property}. This result implies that a
multipartite PPT state of rank four is entangled if and only if its range contains a product vector, see Theorem
\ref{thm:PPTrankfour=noPRODsate}. In the next section we show that this condition can be easily tested. We also show that any
three-qubit PPTES of rank four is strongly extreme in Proposition
\ref{pp:stronglyextreme}.

 \bpp
 \label{pp:PPTESrankAB=4}
Suppose $\r$ is a three-qubit PPTES of rank four. Then
$r(\r_{ij})=4$ for all $1\le i<j\le3$.
 \epp
 \bpf
Suppose that some $r_{ij}:=r(\r_{ij})<4$, say $r_{12}<4$. Then
Proposition \ref{prop:PPTMxNrankN} (i) implies that $\r$, viewed
as a bipartite state for the partition $A_1:A_2A_3$, is separable. Thus, we have
$\r=\sum^n_{i=1}\proj{a_i}_{A_1}\ox(\r^{(i)})_{A_2A_3}$,
where $n>1$ and the $\ket{a_i}$ are pairwise nonparallel. If
$n=2$ then $\r$ is $A_1$-reducible, and Lemma \ref{le:reducible}
(i) gives a contradiction. Thus $n>2$. Since $\r$ is entangled,
we may assume that $\r^{(1)}$ is entangled. Since $r_{12}<4$
and $n>1$, we must have $r_{12}=3$ and so $\cR(\r_{12})$ is spanned by $\ket{a_1,0}$, $\ket{a_1,1}$ and $\ket{a_2,b}$. By
applying an ILO we can assume that $\ket{a_1}=\ket{0}$,
$\ket{a_2}=\ket{1}$ and $\ket{b}=\ket{0}$. Since the $\ket{a_i}$
are pairwise nonparallel, we have $(\r^{(i)})_{A_2}\propto\proj{0}$ for $i>1$ and so
 \bea
 \r =\proj{0}_{A_1}\ox(\r^{(1)})_{A_2A_3} + \sum^n_{i=2} \proj{a_i,0}_{A_1A_2}\ox(\r^{(i)})_{A_3}.
 \eea
Evidently $r(\r^{(1)})<4$, and so we have three cases. We shall
prove that each of them leads to a contradiction.

Case 1: $r(\r^{(1)})=1$. Then \cite[Theorem 5]{cd11JPA} implies
that $\r_{23}$ is NPT and we have a contradiction.

Case 2: $r(\r^{(1)})=3$. For $i>1$, $\ket{0}$ and $\ket{a_i}$ are
linearly independent and so $(\r^{(i)})_{A_3}$ has rank one.
By the same argument, any two states $(\r^{(i)})_{A_3}$ with
$i>1$ are linearly dependent. It follows that $\r$ is
$A_1$-reducible and Lemma \ref{le:reducible} (i) gives a
contradiction.

Case 3: $r(\r^{(1)})=2$. Since $\r^{(1)}$ is a two-qubit entangled
state, we have $\r^{(1)}=\proj{\ps}+\proj{\b,\g}$. Since $\r_{23}$
is PPT, \cite[Theorem 5]{cd11JPA} implies that $\ket{0}$ and
$\ket{\b}$ are linearly independent. Let $\s=\sum^n_{i=2}
\proj{a_i}_{A_1}\ox(\r^{(i)})_{A_3}$. As $r(\r)=4$, we must have
$r(\s)>1$. Hence, $\cR(\r)$ has a basis consisting of vectors
$\ket{0,\ps}$, $\ket{0,\b,\g}$, $\ket{1,0,c_1}$ and
$\ket{a_i,0,c_2}$ for some $i>1$. Suppose that $r(\s)>2$. Then,
after applying an ILO, we may assume that $\cR(\r)$ contains a
vector of the form
$(\ket{0}+\ket{1})\ket{0,x}=y_1\ket{0,\ps}+y_2\ket{0,\b,\g}+y_3\ket{1,0,c_1}+y_4\ket{a_i,0,c_2}$, with $y_1\ne0$ or $y_2\ne0$. It follows that there is a nonzero vector
$\ket{0,x'}\in\cR(\r^{(1)})$. Since $\ket{0}$ and $\ket{\b}$
are linearly independent, we have a contradiction. We conclude
that $r(\s)=2$. Lemma \ref{le:PPTranktwo} implies that
$\s=\proj{a_1,c_1}+\proj{a_2,c_2}$. Next, Lemma
\ref{le:reducible} (ii) implies that $\ket{a_i}$, $i=0,1,2$ are
pairwise linearly independent. By a similar argument one can show
that $\ket{c_1}$ and $\ket{c_2}$ are linearly independent. Thus
$r_{13}=4$ and \cite[Theorem 5]{cd11JPA} implies that the
bipartite state $\r_{2:13}$ is NPT.
Hence, we obtain yet another contradiction.

Since we examined all three cases, the proof is completed.
 \epf

Given a basis $\ket{\ph_i}$, $i=1,\ldots,m$, of some Hilbert
space, there is a unique basis, say $\ket{\ps_i}$,
$i=1,\ldots,m$, such that $\braket{\ph_i}{\ps_j}=\d_{ij}$
for all $i,j$. We say that two such bases are {\em reciprocal}
to each other.

  \bpp
 \label{pp:3qubitPPTES,property}
Let $\r$ be a three-qubit PPTES of rank four. Then

 (i) $r(\r^{\G_S})=4$ for all $S\sue\{1,2,3\}$.

 (ii) $\cR(\r)$ is a CES.

 (iii) We have
 \bea \label{eq:Form1-rho}
\r=\sum^4_{i=1}\proj{a_i}_{A_1}\ox\proj{\ps_i}_{A_2A_3},
 \eea
where any two $\ket{a_i}$ are linearly independent, the
$\ket{\ps_i}$ are linearly independent and each of them is
entangled.

Let $\{\ket{\ps'_i}\}$ be the basis of $\cH_2\ox\cH_3$
reciprocal to $\{\ket{\ps_i}\}$.

 (iv) $\cR(\r)$ and $\ker(\r)$ contain each exactly four
bipartite product vectors for the partition $A_1:A_2A_3$,
namely the $\ket{a_i}\ox\ket{\ps_i}$ and the
$(\ket{a_i^\perp}\ox\ket{\ps'_i})$, respectively.

 (v) Two normalized three-qubit PPTES of rank four having the
same range are equal.
 \epp
 \bpf
By Lemma \ref{le:reducible}, $\r$ is irreducible.
By Propositions \ref{prop:PPTMxNrankN} and
\ref{pp:PPTESrankAB=4}, we know that $\r$ can be written as in
Eq. \eqref{eq:Form1-rho} with linearly independent $\ket{\ps_i}$.

(i) It is immediate from \eqref{eq:Form1-rho} that the assertion
holds if $S=\{1\}$ or $S=\{2,3\}$. Since we can permute the
qubits, it holds in general.

(ii) Suppose $\cR(\r)$ is not a CES. Then $\cR(\r)$ contains a
bipartite product vector for the partition $A_1:A_2A_3$. As
$\cR(\r)$ is spanned by the vectors $\ket{a_i}\ox\ket{\ps_i}$
with linearly independent $\ket{\ps_i}$, at least two vectors $\ket{a_i}$ must be parallel, say $\ket{a_1}\propto\ket{a_2}$.
Since the 2-dimensional subspace spanned by $\ket{\ps_1}$ and $\ket{\ps_2}$ contains a product vector, we may assume that $\ket{\ps_1}=\ket{\b,\g}$. Thus
 \bea \label{eq:Form2-rho}
 \r =
 \proj{\a,\b,\g}
 +
 \sum^4_{i=2}\proj{a_i}_{A_1}\ox\proj{\ps_i}_{A_2A_3},
 \eea
where we have set $\ket{\a}=\ket{a_1}$.
By applying a similar argument to, say, the partition
$A_2:A_1A_3$, we obtain a formula similar to
Eq. \eqref{eq:Form2-rho} in which the first term on the right
hand side is replaced by a scalar multiple of $\proj{\b,\a,\g}$.
This observation implies that for any $S\sue\{1,2,3\}$ we have
$\ket{\a,\b,\g}^S\in\cR(\r^{\G_S})$, where $\ket{\a,\b,\g}^S$
is the product vector obtained from $\ket{\a,\b,\g}$ by replacing
its $i$th factor with the complex conjugate whenever $i\in S$.
Consequently, the operator $\s:=\r-t\proj{a_1,\a,\b}$ is a
PPT state for small $t>0$. We can choose $t>0$ such that
for some $S\sue\{1,2,3\}$, the state $\s^{\G_S}$ is still PPT but
has rank three. Then Theorem \ref{thm:PPTrankthree} implies that
$\s^{\G_S}$ is separable, and so is $\r$. This contradiction
proves that $\cR(\r)$ must be a CES.

(iii) Since $\ket{a_i,\ps_i}\in\cR(\r)$, (ii) implies that
each $\ket{\ps_i}$ is entangled. If say
$\ket{a_1}\propto\ket{a_2}$, then the fact that the 2-dimensional
subspace spanned by $\ket{\ps_1}$ and $\ket{\ps_2}$ contains a
product vector would contradict (ii).

(iv) This assertion follows immediately from (iii).

(v) By (i) and Proposition \ref{prop:PPTMxNrankN}, we shall use the
following analog of Eq. \eqref{eq:Form1-rho}:
 \bea \label{eq:Form3-rho}
\r=\sum^4_{i=1}\proj{b_i}_{A_2}\ox\proj{\ph_i}_{A_1A_3}.
 \eea
Let $\s$ be also a PPTES with $\cR(\s)=\cR(\r)$. Then part (iv)
implies that
 \bea
 \label{ea:rangePPTESsigma}
 \s
 =
 \sum^4_{i=1} p_i \proj{a_i}_{A_1}\ox\proj{\ps_i}_{A_2A_3}
 =
 \sum^4_{i=1} q_i \proj{b_i}_{A_2}\ox\proj{\ph_i}_{A_1A_3}
 \eea
where $p_i,q_i>0$. Let $A$ and $B$ be the $8\times4$ matrices whose
columns are the components of the vectors $\ket{a_i}\ox\ket{\ps_i}$
and $\ket{b_i}\ox\ket{\ph_i}$, respectively.

The Eqs. \eqref{eq:Form1-rho} and \eqref{eq:Form3-rho} imply that
$AA^\dag=BB^\dag$. Similarly, we have $APA^\dag=BQB^\dag$, where
$P=\diag(p_1,\ldots,p_4)$ and $Q=\diag(q_1,\ldots,q_4)$. We can
choose an invertible matrix $V$ such that $VA=[I_4~0]^\dag$. By
writing $VB=\left[\begin{array}{c}X\\Y\end{array}\right]$, with $X$
and $Y$ square matrices, the equation $AA^\dag=BB^\dag$ implies that
$Y=0$ and $X$ is unitary. From the equation $VAPA^\dag
V^\dag=VBQB^\dag V^\dag$ we deduce that $P=XQX^\dag$. Thus, the
diagonal matrices $P$ and $Q$ must have the same spectrum, and by
permuting the $\ket{a_i}\ox\ket{\ps_i}$ we may assume that $P=Q$.
So, we have $P=XPX^\dag$, i.e., $PX=XP$. Suppose that one of the
eigenvalues $p_i$ of $P$, say $p_1$, is simple. Then $X$ breaks into
a direct sum of two matrices, the first one being just $1\times1$
matrix. This implies that $\ket{a_1,\ps_1}\propto\ket{b_1,\ph_1}$.
Therefore $\ket{\ps_1}$ must be a product vector, which contradicts
part (iii). Hence, each $p_i$ must have multiplicity at least two.
Suppose now that $P$ is not a scalar matrix, say $p_1=p_2\ne
p_3=p_4$. Then $X$ breaks into a direct sum $X=X_1\oplus X_2$ of two
$2\times2$ matrices. This implies that the states $\Psi=\sum_{i=1}^2
\proj{a_i}_{A_1}\ox\proj{\ps_i}_{A_2A_3}$ and $\Phi=\sum_{i=1}^2
\proj{b_i}_{A_2}\ox\proj{\ph_i}_{A_1A_3}$ are equal. So the state
$\bra{a_1^\perp}\Psi\ket{a_1^\perp}\propto\proj{\ps_2}$ is
separable, which is a contradiction with (iii). We conclude that $P$
must be a scalar matrix, and so $\s\propto\r$. This concludes the
proof.
 \epf

 \bex
 {\rm
To illustrate Proposition \ref{pp:3qubitPPTES,property}, we consider the three-qubit PPTES $\r$ constructed from the well 
known \cite[Eq. (22)]{DiV03} three-qubit UPB 
 \bea
\ket{000},~\ket{+,1,-},~\ket{1,-,+},~\ket{-,+,1},
 \eea
where $\ket{\pm}=(\ket{0}\pm\ket{1})/2$. The state $\r$ is the normalized projector whose kernel is spanned by the four product vectors of this UPB. Explicitly, we have
 \bea
 \label{ea:divincenzo}
 \r &=&
 \proj{+,\ps_1} + \proj{-,\ps_2} +
 \proj{0,\ps_3} + \proj{1,\ps_4}.
 \eea
where
 \bea
 \ket{\ps_1} &=& \frac{1}{\sqrt{6}}(2\ket{01}+\ket{10}+\ket{11}),
\\
 \ket{\ps_2} &=& \frac{1}{\sqrt{6}}(-\ket{01}-2\ket{10}+\ket{11}),
 \\
 \ket{\ps_3} &=& \frac{1}{\sqrt{3}}(-\ket{01}+\ket{10}+\ket{11}),
  \\
 \ket{\ps_4} &=&
 \frac{1}{\sqrt{12}}(3\ket{00}-\ket{01}+\ket{10}+\ket{11}).
 \eea
Note that Eq. \eqref{ea:divincenzo} is exactly the decomposition 
Eq. \eqref{eq:Form1-rho} in part (iii) of Proposition
\ref{pp:3qubitPPTES,property}. Other assertions of that proposition can be easily verified.
  }
 \eex

We recall from \cite{cd11JMP} that the two-qutrit PPT state of rank
four is entangled if and only if its range is a CES. The analogous
result is valid for three-qubit states, see Proposition
\ref{pp:3qubitPPTES,property} (ii). These observations and Theorem
\ref{thm:PPTrankthree}, \ref{thm:PPTrankfour} imply the following
important result.
 \bt
 \label{thm:PPTrankfour=noPRODsate}
Let $\r$ be a multipartite PPT state of rank $r\le4$. If $r<4$
then $\r$ is separable. If $r=4$ then $\r$ is entangled if
and only if $\cR(\r)$ is a CES.
 \et

In the next section we show that the CES condition can be easily
checked. This is based on the so-called Chow form of Segre
varieties. It has been shown in \cite[Theorem 2]{agk10} that any
normalized three-qubit PPTES $\r$ such that
$r(\r)=r(\r^{\G_1})=r(\r^{\G_2})=4$ is an extreme point of the set
of all PPT states. We conclude this section with the following
observation.

 \bpp
 \label{pp:stronglyextreme}
Any three-qubit PPTES of rank four is strongly extreme.
 \epp
 \bpf
Let $\r$ be a PPTES of rank four. By Theorem \ref{thm:PPTrankfour},
$\r$ is supported on a $3\ox3$ or $2\ox2\ox2$ subsystem. By Theorem
\ref{thm:3x3PPTstates} (iv), the assertion holds in the former case.
For the latter case, let us consider the a PPT state $\s$ such that
$\cR(\s)\sue\cR(\r)$. By Proposition \ref{pp:3qubitPPTES,property}
(ii), $\s$ must be entangled. Lemma \ref{le:PPTranktwo} and Theorem
\ref{thm:PPTrankthree} imply that $r(\s)=4$. Then Proposition
\ref{pp:3qubitPPTES,property} (v) implies that $\r\propto\s$. This
completes the proof.
 \epf
By Definition \ref{def:StronglyExtreme}, any three-qubit PPTES of
rank four is also extreme.

Note that neither of Theorem \ref{thm:PPTrankfour=noPRODsate} and
Proposition \ref{pp:stronglyextreme} holds for PPTES of rank five.
For example, let $\r$ be a two-qutrit PPTES of rank four. Then
$\s=\r+e\proj{00}$ is a two-qutrit PPTES of rank five when $e>0$ is
sufficiently small. Evidently $\cR(\s)$ is not a CES. Furthermore,
$\s$ is not extreme and thus not strongly extreme by Definition
\ref{def:StronglyExtreme}.

\section{\label{sec:chowform} Chow forms of some Segre varieties}

In this section we show how one can test whether the CES
condition of Theorem \ref{thm:PPTrankfour=noPRODsate} is
satisfied. Let us first recall some facts from Algebraic Geometry that we need.

Let $X\subseteq\cP$ be an irreducible projective variety
embedded in a complex projective space $\cP$.
If $L\subseteq\cP$ is a linear subspace of dimension
$\dim\cP-\dim X$ then it is well known that $L\cap X\ne\es$.
On the other hand, the set of all linear subspaces $L$
of dimension $\dim\cP-\dim X-1$ which meet $X$ is a closed
irreducible hypersurface in the corresponding Grassmannian.
This hypersurface is known as the {\em associated hypersurface}
of $X$. It is defined by an equation $F=0$, where $F$
is an irreducible homogeneous polynomial in the Pl\"ucker
coordinates of $L$ known as the {\em Chow form} of $X$.
It is known that the degree of the polynomial $F$ is eaqual
to the degree of the variety $X$.
For more details about the associated hypersurface, see
\cite[Chapter 3]{gkz94}.

If $V$ is a subspace of $\cH$ of dimension
$d-\sum_i(d_i-1)$ then $V$ must contain a product vector.
This is not true for subspaces of smaller dimension.
The case when
\bea \label{eq:UslovDim}
\dim V=\d_1:=d-1-\sum_{i=1}^n (d_i-1)
\eea
is of interest because the set of all vector subspaces of this
dimension which contain a product vector form an irreducible
hypersurface in the corresponding Grassmannian.
This is the associated hypersurface of the Segre variety
$\Sigma=\cP^{d_1-1}\times\cdots\times\cP^{d_n-1}
\subseteq\cP^{d-1}$ canonically embedded in the projective
space $\cP^{d-1}$ of $\cH$. It is defined by a single homogeneous
polynomial equation $F(p_V)=0$ in the Pl\"ucker coordinates
\bea \label{eq:Pluecker}
p_V=\{p_{i_1i_2\ldots i_{\d_1}}:0<i_1<i_2<\cdots<i_{\d_1}\le d\}
\eea
of $V$. The degree of $F$ is equal to the degree of $\S$.

Let us give a few examples of Chow forms $F$ of Segre varieties
$\Sigma$ for bipartite quantum systems $M\ox N$.
In all cases $F$ will be given as a determinant of a square
matrix of order $\d=\binom{M+N-2}{M-1}$.
The Pl\"{u}cker coordinates of $V$ have to be computed in a
product basis, say $\ket{a_i,b_j}$
$(i=1,\ldots,M;~j=1,\ldots,N)$.
(See the Example \ref{PrimerChowForm}.)
It is not required that this basis be orthogonal.
The basis has to be ordered, and we shall
use always the increasing lex ordering:
$\ket{a_1,b_1},\ket{a_1,b_2},\ldots,\ket{a_1,b_N},\ket{a_2,b_1},
\ldots,\ket{a_M,b_N}$.
We start with the cases where $N=2$.

For $(M,N)=(2,2)$ we have
\bea \label{eq:Fza2x2}
F=\left| \begin{array}{cc}
p_{1} & p_{2} \\
p_{3} & p_{4}
\end{array} \right|.
\eea

For $(M,N)=(3,2)$ we have
\bea \label{eq:Fza3x2}
F=\left| \begin{array}{ccc}
p_{13} & p_{14}+p_{23} & p_{24} \\
p_{15} & p_{16}+p_{25} & p_{26} \\
p_{35} & p_{36}+p_{45} & p_{46}
\end{array} \right|.
\eea

For $(M,N)=(4,2)$ we have
\bea \label{eq:Fza4x2}
F=\left| \begin{array}{cccc}
p_{135}  & p_{136}+p_{145}+p_{235}
& p_{146}+p_{236}+p_{245} & p_{246} \\
p_{137} & p_{138}+p_{147}+p_{237}
& p_{148}+p_{238}+p_{247} & p_{248} \\
p_{157} & p_{158}+p_{167}+p_{257}
& p_{168}+p_{258}+p_{267} & p_{268} \\
p_{357} & p_{358}+p_{367}+p_{457}
& p_{368}+p_{458}+p_{467} & p_{468}
\end{array} \right|.
\eea

This can be generalized to any pair $(M,2)$.

\bpp \label{pp:ChowMx2}
The Chow form $F$ of the Segre variety
$\cP^{M-1}\times\cP^1$ embedded canonically in the
projective space $\cP^{2M-1}$ is the determinant of
the $M\times M$ matrix $B=[b_{ij}]$, whose entry $b_{ij}$
is the sum
\bea
b_{ij}=\sum_{P_{ij}} p_{k_1,k_2,\ldots,k_{M-1}}
\eea
of Pl\"ucker coordinates $p_{k_1,k_2,\ldots,k_{M-1}}$
taken over the set $P_{ij}$ of all $\binom{M-1}{j-1}$ sequences
$(k_1,k_2,\ldots,k_{M-1})$ obtained from the sequence
$(1,3,\ldots,\hat{2(M-i)+1},\ldots,2M-1)$ by incrementing by
one exactly $j-1$ of its terms.
(The circumflex indicates that the term $2(M-i)+1$ should be
omitted.)
\epp

The proof of this proposition is given in Appendix I.

Let us point out that if one has a formula for the Chow form
$F=F_{M,N}$ of the Segre variety $\S_{M,N}$ embedded in the
projective space $\cP_{A,B}$, one can easily find the Chow
form $F'=F_{N,M}$ of the Segre variety $\S_{N,M}$ embedded in the
projective space $\cP_{B,A}$. We shall illustrate this
procedure in the case $(M,N)=(3,2)$.
The form $F=F_{3,2}$, given by Eq. \eqref{eq:Fza3x2}, is a
cubic polynomial in the Pl\"{u}cker coordinates $p_{ij}$
computed in the ordered basis
$\ket{00},\ket{01},\ket{10},\ket{11},\ket{20},\ket{21}$
of the Hilbert space $\cH=\cH_A\ox\cH_B$. Note that this basis
consists of bases of the subspaces $\ket{0}\ox\cH_B$,
$\ket{1}\ox\cH_B$, $\ket{2}\ox\cH_B$ written in that order.
To compute the form $F'$, we shall use the ordered basis
$\ket{00},\ket{10},\ket{20},\ket{01},\ket{11},\ket{21}$,
where the first [last] three vectors form a basis of the
subspace $\cH_A\ox\ket{0}$ $[\cH_A\ox\ket{1}]$. Note that
the permutation
\bea
\pi=\left( \begin{array}{cccccc}
1&2&3&4&5&6\\1&4&2&5&3&6 \end{array} \right)
\eea
transforms the first basis into the second. The Chow form
$F_{2,3}$ is obtained from $F_{3,2}$ by replacing each
Pl\"{u}cker coordinate $p_{ij}$ with $p_{\pi(i)\pi(j)}$.
Thus for $(M,N)=(2,3)$ we have
\bea \label{eq:Fza2x3}
F_{2,3}=\left| \begin{array}{ccc}
p_{12} & p_{15}-p_{24} & p_{45} \\
p_{13} & p_{16}-p_{34} & p_{46} \\
p_{23} & p_{26}-p_{35} & p_{56}
\end{array} \right|,
\eea
where we used the fact that $p_{ji}=-p_{ij}$.
Let us compare this result with the one given in
\cite[Example 23]{cd11JPA}. By using the quadratic Pl\"{u}cker
relations $p_{12}p_{34}-p_{13}p_{24}+p_{14}p_{23}=0$,
$p_{14}p_{56}-p_{15}p_{46}+p_{16}p_{45}=0$ and
$p_{24}p_{56}-p_{25}p_{46}+p_{26}p_{45}=0$, one can easily
verify that the polynomial on the left hand side of the
equation on top of p. 16 in \cite{cd11JPA} is equal to
$-F_{2,3}$. Since the Chow form is determined only up to a
scalar factor, the two results agree.

The most important case for us is $(M,N)=(3,3)$.
In that case we have
\bea \label{eq:Fza3x3}
F=\left| \begin{array}{cccrrr}
p_{1245} & p_{1346} & p_{2356}
& \begin{array}{r} p_{1246}+p_{1345} \end{array}
& \begin{array}{r} p_{1256}+p_{2345} \end{array}
& \begin{array}{r} p_{1356}+p_{2346} \end{array} \\
p_{1278} & p_{1379} & p_{2389}
& \begin{array}{r} p_{1279}+p_{1378} \end{array}
& \begin{array}{r} p_{1289}+p_{2378} \end{array}
& \begin{array}{r} p_{1389}+p_{2379} \end{array} \\
p_{4578} & p_{4679} & p_{5689}
& \begin{array}{r} p_{4579}+p_{4678} \end{array}
& \begin{array}{r} p_{4589}+p_{5678} \end{array}
& \begin{array}{r} p_{4689}+p_{5679} \end{array} \\
p_{1248}-p_{1257} & p_{1349}-p_{1367} & p_{2359}-p_{2368} &
\begin{array}{r}
p_{1249}-p_{1267} \\ +p_{1348}-p_{1357}
\end{array} &
\begin{array}{r}
p_{1259}-p_{1268} \\ +p_{2348}-p_{2357}
\end{array} &
\begin{array}{r}
p_{1359}-p_{1368} \\ +p_{2349}-p_{2367}
\end{array} \\
p_{1458}-p_{2457} & p_{1469}-p_{3467} & p_{2569}-p_{3568} &
\begin{array}{r}
p_{1459}-p_{2467} \\ +p_{1468}-p_{3457}
\end{array} &
\begin{array}{r}
p_{1568}-p_{2567} \\ +p_{2459}-p_{3458}
\end{array} &
\begin{array}{r}
p_{1569}-p_{3468} \\ +p_{2469}-p_{3567}
\end{array} \\
p_{1578}-p_{2478} & p_{1679}-p_{3479} & p_{2689}-p_{3589} &
\begin{array}{r}
p_{1579}-p_{2479} \\ +p_{1678}-p_{3478}
\end{array} &
\begin{array}{r}
p_{1589}-p_{2489} \\ +p_{2678}-p_{3578}
\end{array} &
\begin{array}{r}
p_{1689}-p_{3489} \\ +p_{2679}-p_{3579}
\end{array}
\end{array} \right|.
\eea

The Chow form of the Segre variety of the system
$2\ox2\ox2$ of three qubits is given explicitly in
\cite[Proposition 4.10]{bs02}. Some misprints in the
determinant above that proposition have been corrected
in \cite[Example 22, Eq. (23)]{dss06}. We point out that
the determinantal formula in these references is
written in terms of the dual Pl\"ucker coordinates.
When translated into ordinary Pl\"ucker coordinates,
the Chow form is given by the following determinant:

\bea \label{eq:Fza2x2x2}
F = \left| \begin{array}{ccccrr}
p_{1235} & p_{1237} & p_{1567} & p_{3567}
& \begin{array}{r} p_{1257}-p_{1356} \end{array}
& \begin{array}{r} p_{1367}-p_{2357} \end{array} \\
p_{1246} & p_{1248} & p_{2568} & p_{4568}
& \begin{array}{r} p_{1268}-p_{2456} \end{array}
& \begin{array}{r} p_{1468}-p_{2458} \end{array} \\
p_{1345} & p_{1347} & p_{1578} & p_{3578}
& \begin{array}{r} p_{1358}-p_{1457} \end{array}
& \begin{array}{r} p_{1378}-p_{3457} \end{array} \\
p_{2346} & p_{2348} & p_{2678} & p_{4678}
& \begin{array}{r} p_{2368}-p_{2467} \end{array}
& \begin{array}{r} p_{2478}-p_{3468} \end{array} \\
p_{1236}+p_{1245} & p_{1238}+p_{1247} &
p_{1568}+p_{2567} & p_{3568}+p_{4567}
& \begin{array}{r}
p_{1258}-p_{1456} \\ +p_{1267}-p_{2356}
\end{array} &
\begin{array}{r}
p_{1368}-p_{2358} \\ +p_{1467}-p_{2457}
\end{array} \\
p_{1346}+p_{2345} & p_{1348}+p_{2347} &
p_{1678}+p_{2578} & p_{3678}+p_{4578}
& \begin{array}{r}
p_{1368}-p_{1467} \\ +p_{2358}-p_{2457}
\end{array} &
\begin{array}{r}
p_{1478}-p_{3458} \\ +p_{2378}-p_{3467}
\end{array} \\
\end{array} \right|.
\eea

The problem of deciding whether a bipartite PPT state $\r$ of
rank four is separable has been solved in
\cite[Theorem 22]{cd11JPA}. The answer is affirmative if and only if $\cR(\r)$ contains at least one product vector.
The proof easily reduces to the case when $\r$ is a
$3\times3$ state. In that case we can improve the mentioned
theorem. By Theorem \ref{thm:PPTrankfour=noPRODsate}, the analogous result is valid for the PPT three-qubit states of rank four. Thus we have the following result.

\bt \label{thm:3x3stanje4}
A $3\times3$ or $2\times2\times2$ state $\r$ of rank four is separable if and only if $\r$ is PPT and the Pl\"ucker coordinates of $\cR(\r)$ satisfy the equation $F=0$ where $F$ is the Chow form \eqref{eq:Fza3x3} or \eqref{eq:Fza2x2x2}, 
respectively.
\et

We illustrate this theorem by the following example.

\bex \label{PrimerChowForm}
$(M=N=3)$ {\rm
Let $\r=\sum_i\proj{\psi_i}$ where
$\ket{\psi_1}=\ket{00}+a\ket{11}$,
$\ket{\psi_2}=a\ket{01}+\ket{10}+b\ket{21}$,
$\ket{\psi_3}=\ket{11}+b\ket{20}+\ket{22}$,
$\ket{\psi_4}=\ket{12}+\ket{21}$
and $a,b$ are complex parameters.
It is easy to verify that this is always a PPT state of rank
four. The range of $\r$ is spanned by the rows of the matrix
\bea
R=\left[ \begin{array}{ccccccccc}
1 & 0 & 0 & 0 & a & 0 & 0 & 0 & 0 \\
0 & a & 0 & 1 & 0 & 0 & 0 & b & 0 \\
0 & 0 & 0 & 0 & 1 & 0 & b & 0 & 1 \\
0 & 0 & 0 & 0 & 0 & 1 & 0 & 1 & 0
 \end{array} \right].
\eea
It is now easy to compute the Pl\"ucker coordinates
$p_{ijkl}$ of $\cR(\r)$. Recall that $p_{ijkl}$ is the
determinant of the full submatrix of $R$ in columns
$i,j,k,l$. We find that
\bea
&& p _{1456}=p_{1458}=1, \notag \\
&& p _{1469}=p_{1489}=-1, \notag \\
&& p _{1256}=p_{1258}=p_{4569}=p_{4589}=a, \notag \\
&& p _{1269}=p_{1289}=-a, \notag \\
&& p _{1478}=p_{1568}=p_{1689}=b, \notag \\
&& p _{1467}=-b, \notag \\
&& p_{2569}=p_{2589}=a^2, \notag \\
&& p _{1278}=p_{4567}=p_{5689}=ab, \notag \\
&& p _{1267}=p_{4578}=-ab, \notag \\
&& p _{1678}=-b^2, \notag \\
&& p_{2567}=a^2b, \notag \\
&& p _{2578}=-a^2b, \notag \\
&& p_{5678}=-ab^2, \notag
\eea
and all other are 0.
By plugging in these values into Eq. \eqref{eq:Fza3x3},
we obtain that $F=-a^4b^4$.
Hence, $\r$ is separable if and only if $ab=0$.
\hfill $\square$ }
\eex

\acknowledgments

We thank Marco Piani for pointing out Ref. \cite{aaa12}. The first
author was mainly supported by MITACS and NSERC. The CQT is funded
by the Singapore MoE and the NRF as part of the Research Centres of
Excellence programme. The second author was supported in part by an
NSERC Discovery Grant. He is grateful to B. Sturmfels for providing
references \cite{gkz94,bs02,dss06}.

\section*{\label{sec:Dokaz}
Appendix I: Proof of Proposition \ref{pp:ChowMx2}}

We shall now derive Proposition \ref{pp:ChowMx2} from
Theorem 3.19 and Proposition 3.21 of \cite[Chapter 14]{gkz94}.

Let us first recall these results and introduce the necessary
notation.
For convenience we set $[n]=\{1,2,\ldots,n\}$ for any positive
integer $n$. We consider the bipartite system $M\ox N$.
We are interested in subspaces $V$ of $\cH=\cH_A\ox\cH_B$ of
dimension $(M-1)(N-1)$. Such $V$ is defined by a system of
$M+N-1$ independent linear homogeneous equations
$\sum_{i,j,k} a_{ijk}\x_{ik}=0$, $j\in[M+N-1]$. We assume here
that the vectors $\ket{x}\in V$ are given as linear combinations
of standard basis vectors
$\ket{x}=\sum_{i,k} \x_{ik}\ket{i-1}\ox\ket{k-1}$.
The coefficients $a_{ijk}$ form a 3-dimensional matrix
$A=[a_{ijk}]$, where $i\in[M]$, $j\in[M+N-1]$ and $k\in[N]$.
We can rearrange the entries of $A$ to obtain an ordinary
$(M+N-1)\times MN$ matrix $A^{13}$. First, let us note that any
$r\in[MN]$ can be uniquely written as $r=(i_r-1)N+k_r$ with
$i_r\in[M]$ and $k_r\in[N]$. We define $\tilde{r}=(i_r,k_r)$
for $r\in[MN]$. Then we define the $(j,r)$th entry of $A^{13}$
to be $a_{i_r,j,k_r}$ where $\tilde{r}=(i_r,k_r)$.

For any sequence $r_1,r_2,\ldots,r_{M+N-1}$ in $[MN]$, we denote
by $q_{r_1,r_2,\ldots,r_{M+N-1}}$ the determinant of the square
matrix formed from the columns of $A^{13}$ writen in the indicated order: first column $r_1$, then column $r_2$, etc.
The {\em dual Pl\"{u}cker coordinates} of $V$ are the
determinants $q_{r_1,r_2,\ldots,r_{M+N-1}}$ with
$r_1<r_2<\cdots<r_{M+N-1}$. There is a simple relation between
the dual and ordinary Pl\"{u}cker coordinates of $V$,
see Eq. (1.6) in \cite[Chapter 3]{gkz94}. In our case it has
the following form
\bea \label{eq:dual/ord}
q_{r_1,r_2,\ldots,r_{M+N-1}}=\varepsilon
p_{r'_1,r'_2,\ldots,r'_{(M-1)(N-1)}},
\eea
where the indexes $r'_1,\ldots,r'_{(M-1)(N-1)}$ are arranged in
increasing order and form the
complement of $\{r_1,\ldots,r_{M+N-1}\}$ in $[MN]$.
The symbol $\varepsilon=\pm1$ is the sign of the permutation
$r'_1,\ldots,r'_{(M-1)(N-1)},r_1,\ldots,r_{M+N-1}$.

Another piece of notation that we need is
\bea \label{def:Delta}
\Delta^{p-1}(m)=\{\a=(\a_1,\ldots,\a_p)\in\bZ_+^p:
\a_1+\cdots+\a_p=m\},
\eea
where $\bZ_+$ is the set of nonnegative integers.

As shown in Theorem 3.19 of \cite[Chapter 14]{gkz94}, the Chow
form $F$ of the Segre variety
$\cP^{M-1}\times\cP^{N-1}$ in $\cP^{MN-1}$ is given by
$F=\det B$, where $B=[b_{\a,\b}]$ is a square matrix of order
$\d$, and $\a\in\Delta^{M-1}(N-1)$ and $\b\in\Delta^{N-1}(M-1)$.
Furthermore, there is also a formula for the matrix entries
$b_{\a,\b}$. To state this formula, we need to
introduce the complete bipartite graph whose vertex set is
the disjoint union of $[M]$ and $[N]$ and the edge set is
the Cartesian product $[M]\times[N]$. Any spanning tree,
$\Omega$, of this graph consists of $M+N-1$ edges.
Moreover, $\Omega$ can be written uniquely as
\bea \label{eq:Omega}
\Omega=\{(1,k_1),\ldots,(M,k_M),(i_2,2),\ldots,(i_N,N)\}.
\eea
We remark that the indexes $k$ and $i$ are not arbitrary.
For instance, since $|\Omega|=M+N-1$ we must have
$k_{i_r}\ne r$ for $r\in\{2,3,\ldots,N\}$.
Then we define $[\Omega]=q_{r_1,r_2,\ldots,r_{M+N-1}}$, where
\bea \label{eq:OmegaIndexes}
r_s=(s-1)N+k_s, \quad s\in[M]; \quad
r_{M+s}=(i_s-1)N+s, \quad s=2,3,\ldots,N.
\eea
For $(\a,\b)\in\Delta^{M-1}(N-1)\times\Delta^{N-1}(M-1)$
with $\a=(\a_1,\ldots,\a_M)$ and $\b=(\b_1,\ldots,\b_N)$,
we shall write $\Omega\;\vdash(\a,\b)$ if
$\a_r=|\{s:i_s=r\}|$ for $r\in[M]$ and
$\b_s=|\{i:k_i=s\}|$ for $s\in\{2,3,\ldots,N\}$.
With this notation, Proposition 3.21 of \cite[Chapter 14]{gkz94} asserts that the entries $b_{\a,\b}$ are given by the formula
\bea \label{eq:b-entries}
b_{\a,\b}=\sum_{\Omega\;\vdash(\a,\b)} [\Omega].
\eea

We can now prove Proposition \ref{pp:ChowMx2}.
By hypothesis we have $N=2$. Thus
$\Delta^{M-1}(N-1)=\{\ve_i:i\in[M]\}$
where the $i$th entry of $\ve_i$ is 1 and all other 0.
Note that $\Delta^{N-1}(M-1)=\{(M-1-s,s):s=0,1,\ldots,M-1\}$.
Let us compute the entry $b_{\a,\b}$ for $\a=\ve_u$ and
$\b=(M-1-v,v)$. Let
$\Omega:=\{(1,k_1),\ldots,(M,k_M),(i_2,2)\}\;\vdash(\a,\b)$.
By using  the definition of the relation ``$\;\vdash\;$'' and the
fact that $\a=\ve_u$, we infer that $i_2=u$.
As mentioned above, we must have $k_{i_2}\ne2$, and so $k_u=1$.
Moreover, the set
$S=\{i:k_i=2\}$ has cardinality $v$. Clearly, for fixed
$u$ and $v$, $\Omega$ is determined uniquely by the set $S$.
Note that $S$ is subject only to the conditions that
$|S|=v$ and $u\notin S$. Consequently, there are exactly
$\binom{M-1}{v-1}$ maximal trees $\Omega$ such that
$\Omega\;\vdash(\a,\b)$. For convenience, we shall write
$[r_1,\ldots,r_{M+1}]=q_{r_1,\ldots,r_{M+1}}$, and
we shall use similar notation for ordinary Pl\"{u}cker
coordinates. Next note that
\bea \notag
[\Omega] &=& [k_1,2+k_2,4+k_3,\ldots,2M-2+k_M,2u] \\
&=& (-1)^{M-u}[k_1,2+k_2,\ldots,2u-4+k_{u-1},
2u-1,2u,2u+k_{u+1},\ldots,2M-2+k_M]. \label{eq:izraz1}
\eea
Passing to the ordinary Pl\"{u}cker coordinates, we obtain
that
\bea
[\Omega]=(-1)^{M-u}\ve [3-k_1,5-k_2,7-k_3,\ldots,
2u-1-k_{u-1},2u+3-k_{u+1},\ldots,2M+1-k_M],
\eea
where $\ve$ is the sign of the permutation
\bea \notag
\s &=& 3-k_1,5-k_2,7-k_3,\ldots,
2u-1-k_{u-1},2u+3-k_{u+1},\ldots,2M+1-k_M, \\
&& k_1,2+k_2,4+k_3,\ldots,2u-4+k_{u-1},2u-1,2u,2u+k_{u+1},\ldots,2M-2+k_M.
\eea
Since $F$ is defined only up to a scalar factor, we can change
the sign of any row or column of $B$ if necessary. Hence we can
ignore the factor $(-1)^{M-u}$ in Eq. \eqref{eq:izraz1} and
replace $\s$ by the permutation
\bea \notag
\s' &=& k_1,3-k_1,2+k_2,5-k_2,4+k_3,7-k_3,\ldots,
2u-4+k_{u-1},2u-1-k_{u-1},2u-1,2u, \\
&& 2u+k_{u+1},2u+3-k_{u+1},\ldots,2M-2+k_M,2M+1-k_M.
\eea
Since the sign of $\s'$ is $(-1)^v$, we may assume that
\bea
[\Omega]=p_{3-k_1,5-k_2,7-k_3,\ldots,
2u-1-k_{u-1},2u+3-k_{u+1},\ldots,2M+1-k_M}.
\eea
We conclude that after suitable permutation of rows and columns,
and multiplying some rows and columns with $-1$, the
matrix $B$ used in this proof becomes equal to the transpose of
the matrix $B$ defined in the proposition.
This completes the proof.

\end{document}